\documentclass[a4paper]{article}
\usepackage[margin=25mm]{geometry}
\usepackage{amsfonts,amssymb,amsmath,amsthm,dsfont}
\usepackage[linesnumbered,algoruled]{algorithm2e}
\usepackage{graphicx}
\usepackage{subcaption}
\graphicspath{{figures/}}
\usepackage{textcomp}
\usepackage{xcolor}
\usepackage{comment}
\usepackage{multirow}
\usepackage{url}
\usepackage{pifont}
\newcommand{\cmark}{\ding{51}}
\newcommand{\xmark}{\ding{55}}
\usepackage{ulem}
\usepackage{natbib}

\providecommand{\keywords}[1]
{
  \small
  \textbf{\textit{Keywords---}} #1
}

\begin{document}

\title{Neural content-aware collaborative filtering \\for cold-start music recommendation\thanks{The work of P. Magron was conducted while he was with IRIT, Université de Toulouse, CNRS, Toulouse, France. This work has received funding from the European Research Council (ERC) under the European Union’s Horizon 2020 research and innovation programme under grant agreement No 681839 (project FACTORY). Experiments presented in this paper were carried out using the Grid'5000 testbed, supported by a scientific interest group hosted by Inria and including CNRS, RENATER and several Universities as well as other organizations (see https://www.grid5000.fr).}}

\date{}
\author{Paul~Magron\thanks{Université de Lorraine, CNRS, Inria, LORIA, F-54000 Nancy, France (e-mail: firstname.lastname@inria.fr).},
        Cédric~Févotte\thanks{IRIT, Université de Toulouse, CNRS, Toulouse, France (e-mail: firstname.lastname@irit.fr).}
}

\maketitle
\begin{abstract}
State-of-the-art music recommender systems are based on collaborative filtering, which builds upon learning similarities between users and songs from the available listening data. These approaches inherently face the cold-start problem, as they cannot recommend novel songs with no listening history. Content-aware recommendation addresses this issue by incorporating content information about the songs on top of collaborative filtering. However, methods falling in this category rely on a shallow user/item interaction that originates from a matrix factorization framework. In this work, we introduce neural content-aware collaborative filtering, a unified framework which alleviates these limits, and extends the recently introduced neural collaborative filtering to its content-aware counterpart. We propose a generative model which leverages deep learning for both extracting content information from low-level acoustic features and for modeling the interaction between users and songs embeddings. The deep content feature extractor can either directly predict the item embedding, or serve as a regularization prior, yielding two variants ({\em strict} and {\em relaxed}) of our model. Experimental results show that the proposed method reaches state-of-the-art results for a cold-start music recommendation task. We notably observe that exploiting deep neural networks for learning refined user/item interactions outperforms approaches using a more simple interaction model in a content-aware framework.
\end{abstract}

\section{Introduction}

Music recommendation consists in predicting users' listening habits in order to suggest them novel tracks that they might enjoy~\citep{Schedl2015}. This task, which is at the core of many commercial platforms, has been extensively investigated, but remains challenging due to the complexity of music and to the lack of explicit and reliable users feedback~\citep{Hu2008}. In particular, leveraging the musical \textit{function} at hand is necessary to perform recommendation that are tailored for a specific usage~\citep{Schedl2017}. More generally, the usage of contextual information~\citep{Pichl2021} is of paramount importance in order to adapt the recommendation to a particular location or event~\citep{Gillhofer2015,Zhiyong2016}. Besides, it has been shown that psychological cues such as personality and emotional response play a major role in musical tastes and users behavior~\citep{Ferwerda2015}. Recommender systems could then benefit from such findings in music psychology research~\citep{Laplante2014,Soleymani2015,Magron2021avd}. Finally, music recommender systems face the \emph{cold-start} problem~\citep{Schein2002}, which is the topic of investigation of this paper: when a new song is added to a music streaming platform, it has no interaction data with the set of users. Consequently, the system cannot properly recommend this novel item to existing users. Even though this problem has fueled an important amount of studies, it is still considered as a major challenge in music recommendation research~\citep{Schedl2018}.

State-of-the-art approaches for music recommendation are based on collaborative filtering~\citep{Hu2008}, a family of techniques which rely solely on users' listening history: the interest of a given user for a given song is predicted using similarities between various user profiles. The users' feedback are most often implicit and in the form of \emph{playcounts}, that is, how many times a given user has listened to a particular song. Collaborative filtering techniques consist in learning user and item embeddings from the data, where these embeddings respectively characterize the users' preferences and the items' attributes. This listening history is however noisy and lacks negative feedback data~\citep{Hu2008,Rendle2009}. To alleviate this problem, weighted matrix factorization (WMF) techniques~\citep{Salakhutdinov2007,Koren2009} process binarized data computed from the raw playcounts, and associate a measure of \emph{confidence} to these binarized feedback. This family of techniques has shown good performance in music recommender systems~\citep{Liang2015}. More recently, deep learning has been leveraged in collaborative filtering techniques with promising results~\citep{Cheng2016,Zhang2017,Liang2018}. While some earlier models rely on shallow architectures~\citep{Salakhutdinov2007a, Wu2016}, deep neural networks (DNNs) are now exploited for learning deep user/item embeddings and interaction models to replace the matrix product in WMF~\citep{Xue2017,He2017,He2018,Chen2019}.

However, these approaches are agnostic to any form of item-related content, which becomes a major issue for new items. Indeed, since collaborative filtering methods only exploit the available user/item interaction data, they are not able to recommend a song without listening history, and therefore face the cold-start problem. This problem has been tackled with content-based methods, which aim to exploit additional information about the items for recommendation~\citep{Yoshii2006,Wang2011,Fang2011}. Deep learning has been used as a tool to learn features from the content that can help the collaborative filtering. In~\citep{Liang2015}, the last hidden layer of an auto-tagging DNN is used as content feature. Conversely, in~\citep{Oord2013,Wang2014}, acoustic features are mapped to the learned item attribute matrix in order to be further used for cold-start recommendation. These are however limited in performance since the user/item embeddings and the deep content feature extractor are learned in two distinct stages. To alleviate this issue, several recent works have proposed to combine these steps into a single-stage approach, where the user/item embeddings and the content feature extractor are jointly learned. Examples of architectures for the deep content feature extractor include stacked denoising auto-encoders~\citep{Wang2015,Li2015}, variational auto-encoders~\citep{Li2017}, recurrent neural networks~\citep{Wang2016rnn} and convolutional networks~\citep{Huan2017,Lee2018}. However, these approaches still rely on a simple matrix product for modeling the interaction between users and items, and therefore do not leverage DNNs for learning more complex interaction models.

In this work, we introduce neural content-aware collaborative filtering (NCACF), a unified framework which overcomes these limits, and extends the recently introduced neural collaborative filtering (NCF)~\citep{He2017} to its content-aware counterpart. We propose a generative model which leverages DNNs for both extracting content information from low-level acoustic features and for learning refined interaction models between user and item embeddings. Two variants of the model are considered. In the {\em strict} variant, the deep content feature extractor directly predicts the item embedding. Conversely, in the {\em relaxed} variant, it serves as a regularization prior for the item embedding. These two approaches allow for more flexibility and for generalizing previous models from the literature. In particular, when the interaction model reduces to a simple product (which is a building block of matrix factorization models), we derive an estimation algorithm which hybridizes closed-form updates using alternating least squares (ALS) for the embeddings and gradient descent (GD) for the deep content feature extractor. We further incorporate the embeddings in the network in order to estimate the whole model jointly using a single GD algorithm. Finally, we use this technique to estimate the more general NCACF model which uses a deep user/item interaction model. To assess the potential of these methods, we conduct experiments on the Million Song Dataset, a publicly available database for music information retrieval tasks. We observe that the proposed NCACF outperforms the baselines using a shallow user/item interaction model, but also recent content-free approaches using a deep interaction model. This reveals its interest for cold-start music recommendation applications, but also for traditional collaborative filtering tasks, where leveraging side-content information is shown beneficial. To summarize, the main contributions of our work are as follows:
\begin{enumerate}
    \item We introduce NCACF, a deep learning-based recommendation system. This model allows for learning refined user/item interactions as well as leveraging content information, which makes it suitable for cold-start recommendation.
    \item By considering several variants of this model's architecture, underlying generative processes, and training strategies, we show that many methods from the recent literature (both content-free and content-aware) can be cast as particular instances of the NCACF unified framework.
    \item We conduct extensive experiments on the largest publicly available music recommendation dataset, where NCACF reaches state-of-the-art performance. Carefully evaluating its many particular cases notably allows to evaluate the potential of several recent methods for cold-start recommendation, which is complementary to the warm-start setting they were initially designed for.
\end{enumerate}

The rest of this paper is structured as follows. Section~\ref{sec:background} presents the work related to our approach. The proposed NCACF model is then introduced in Section~\ref{sec:models}. The experimental protocol is detailed in Section~\ref{sec:protocol} and the results are presented and discussed in Section~\ref{sec:results}. Finally, Section~\ref{sec:conclu} draws some concluding remarks.

\section{Related work}
\label{sec:background}

\begin{table}[t]
    \centering
    \caption{Mathematical notations.}
    \label{tab:notations}
    \begin{tabular}{ll}
\hline\noalign{\smallskip}
    $U$ & Number of users. \\ 
    $I$ & Number of items (= songs). \\
    $K$ & Dimension of the user/item embeddings.\\
    $L$ & Dimension of the content vectors (= acoustic features).\\
    $P$ & Number of layers in the deep content feature extractor. \\
    $Q$ & Number of hidden layers in the deep interaction model. \\
\noalign{\smallskip}\hline\noalign{\smallskip}
     $\mathbf{Y}$ & User/item interaction matrix (raw playcounts).\\  
     $\mathbf{R}$ & Binarized playcount matrix.\\  
     $\mathbf{r}_u$, $\mathbf{r}_i$ & Binarized playcounts for user $u$ and item $i$.\\  
     $\mathbf{C}$ & Confidence matrix.\\    
     $\mathbf{c}_u$, $\mathbf{c}_i$& Confidence for user $u$ and item $i$.\\
     $\mathbf{x}_i$ & Acoustic feature vector for item $i$.\\
     $\mathbf{W}$, $\mathbf{H}$ & User and item embedding matrices.\\
     $\mathbf{w}_u$, $\mathbf{h}_i$ & Embedding vectors for user $u$ and item $i$.\\
\noalign{\smallskip}\hline\noalign{\smallskip}
     $\phi_{\theta}$ & Deep content feature extractor, with parameters $\theta$.\\  
     $\psi_{\gamma}$ & User/item interaction model, with parameters $\gamma$.\\  
\noalign{\smallskip}\hline\noalign{\smallskip}
     $^{-1}$ & Matrix inverse.\\  
     $\odot$ & Element-wise vector or matrix multiplication. \\
     $.^\mathsf{T}$ & Vector or matrix transpose.\\  
     $\mathbf{I}_K$ & Identity matrix of size $K$.\\  
     $\mathbf{1}$ & Vector or matrix whose entries are all equal to $1$.\\  
     $\text{diag}(\mathbf{a})$ & Diagonal matrix whose entries are given by the vector $\mathbf{a}$.\\  
     $||.||$ & Euclidean norm. \\ 
\noalign{\smallskip}\hline
    \end{tabular}
\end{table}

In this section, we present the related work upon which our method builds.
We first describe the baseline WMF model, and then present approaches that incorporate deep content feature extraction in WMF. Finally, we present the collaborative filtering methods using deep learning for user/item interaction modeling.
The notations used throughout this paper are summarized in Table~\ref{tab:notations}.

\subsection{Weighted matrix factorization}
\label{sec:background_wmf}

Let us consider a large data matrix $\mathbf{Y} \in \mathbb{R}^{U \times I}$ representing interactions between a set of $U$ users and $I$ items, where $y_{u,i}$ denotes the interaction between user $u$ and item $i$. Matrix factorization~\citep{Marlin2004,Hu2008} consists in decomposing the data $\mathbf{Y}$ as the product of two low-dimensional factors: a (transposed) user preferences matrix $\mathbf{W} \in \mathbb{R}^{K \times U}$ and an item attributes matrix $\mathbf{H} \in \mathbb{R}^{K \times I}$, such that $\mathbf{Y} \approx \mathbf{W}^{\mathsf{T}} \mathbf{H}$. The rank $K$ of the decomposition is chosen such that $K(U+I) \ll UI$ to ensure dimensionality reduction. More specifically, weighted matrix factorization (WMF)~\citep{Salakhutdinov2008,Koren2009} is a variant of matrix factorization that is appropriate for handling implicit feedback data. In this variant, the factorization is performed on a binarized interaction data matrix $\mathbf{R}$ (that is, $r_{u,i}=1$ if user $u$ has interacted with item $i$ and $0$ otherwise).

The WMF model can be estimated using alternating least square (ALS)~\citep{Hu2008, He2016}, which yields closed-form updates for the user and item factors. Once these have been estimated, recommendation can be performed using the predicted interaction defined as $\hat{r}_{u,i}=\mathbf{w}_u^{\mathsf{T}}\mathbf{h}_i$, and constructing a recommendation list for each user by picking the items with the highest predicted interaction for this user. Even though WMF has shown good performance for recommender systems~\citep{Hu2008,Oord2013}, it only applies to songs for which some listening history is available, hence facing the cold-start problem.

\subsection{Combining WMF and deep content features}
\label{sec:background_wmf_content}

\begin{figure}[t]
    \centering
    \includegraphics[width=.9\columnwidth]{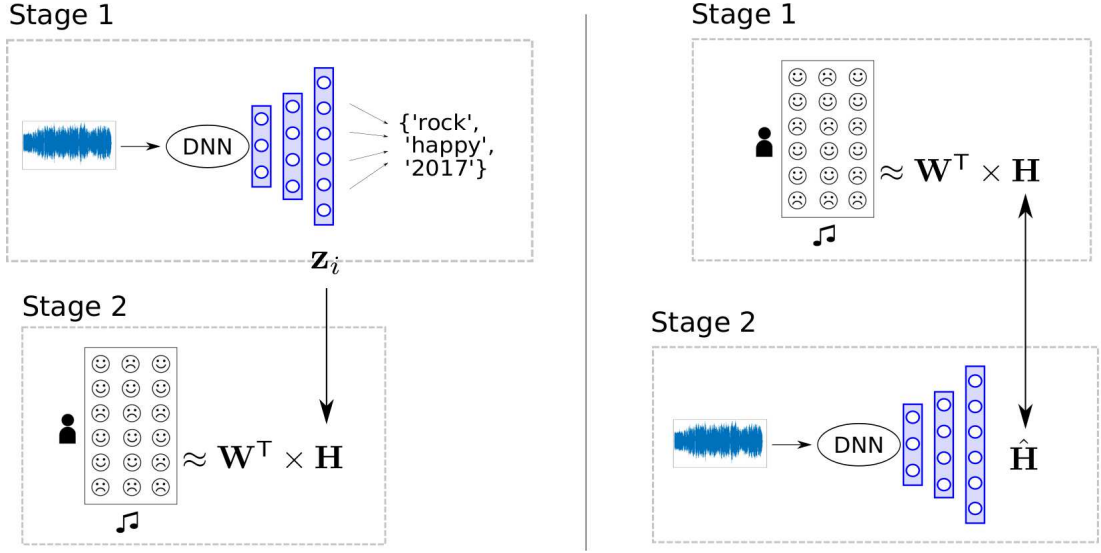}
    \caption{Content-aware WMF methods using deep content features. \cite{Liang2015} (left) extract content features using an auto-tagging DNN and then incorporated it in WMF. Conversely,~\cite{Oord2013} (right) first obtain the WMF decomposition, and then use the learned attribute matrix as target for training the deep content extractor.}
    \label{fig:musrec_2stages}
\end{figure}

To alleviate the aforementioned cold-start problem, several works have proposed to exploit side content information about the songs in a factorization-based framework. While some of these works rely on closed-form models~\citep{Jeunen2020}, e.g., based on matrix co-factorization~\citep{Gouvert2018}, we focus here on approaches that leverage deep learning in conjunction with WMF~\citep{Liang2015,Oord2013}. These are illustrated in Fig.~\ref{fig:musrec_2stages} and described hereafter. 

In~\citep{Liang2015}, the authors propose to incorporate some content information in WMF. More precisely, they consider a prior on the item attribute matrix of the form of $\mathbf{h}_i \approx \boldsymbol{\Phi} \mathbf{z}_i$, where $\mathbf{z}_i$ is the content feature vector, and $\boldsymbol{\Phi}$ linearly maps this vector to the item embedding space. As for the content feature vector $\mathbf{z}_i$, the authors extract the last hidden representation of an auto-tagging DNN that is trained beforehand. This approach allows to alleviate the cold-start problem: for novel items without listening history, binarized playcounts can still be predicted through $\hat{r}_{u,i}=\mathbf{w}_u^{\mathsf{T}} \boldsymbol{\Phi} \mathbf{z}_i$, from which recommendation can be made. However, the quality of the content features is highly dependent on the performance of the auto-tagging network. While these methods have significantly improved in the recent years~\citep{Kim2018,Pons2018}, they are still limited by the noisy nature of the tags, which are mostly user-annotated. Besides, the learned features are not necessarily relevant for the task at hand (i.e., recommendation instead of tagging), since they are neither estimated jointly with the user/item embeddings nor fine-tuned for this task.

\cite{Oord2013} adopt a different strategy in order to leverage WMF for content-based recommendation. First, the factors $\mathbf{W}$ and $\mathbf{H}$ are estimated from the listening data using ALS updates, as mentioned in Section~\ref{sec:background_wmf}. Then, in a second stage, they train a DNN $\phi_{\theta}$ with parameters $\theta$ whose purpose is to map input acoustic features $\mathbf{x}_i$ to the estimated attributes $\mathbf{h}_i$. This DNN can then be used for predicting item attributes for novel songs and therefore perform recommendation in a cold-start setting through $\hat{r}_{u,i}=\mathbf{w}_u^{\mathsf{T}} \phi_{\theta}(\mathbf{x}_i)$. However, this approach strongly relies on the proper estimation of the WMF factors $\mathbf{W}$ and $\mathbf{H}$, which are needed for training $\phi_{\theta}$. Since the authors do not alternate the updates on the factors and the network parameters, the quality of the prior factorization sets a performance limit to this approach.

In a nutshell, these methods are mainly limited by that they operate in two stages, which means that the content feature extractor and the factorization model are not trained jointly. As a result, in~\citep{Liang2015} the extracted features are not necessarily relevant for a recommendation task, and in~\citep{Oord2013} the capacity of the deep content extractor is inherently limited by the prior factorization.

To alleviate these drawbacks, several recent works have proposed to combine these steps into a single-stage approach termed collaborative deep learning (CDL) \citep{Wang2015}. This allows for jointly learning the user/item embeddings and the deep content feature extractor. Various architectures have been proposed for modeling the DNN $\phi_{\theta}$, such as denoising~\citep{Wang2015} or variational~\citep{Li2017} auto-encoders. These approaches propose to estimate the whole model jointly, that is, alternating updates on the factors and the network parameters (this will be described in details in Section~\ref{sec:models_mf}). Predictions are finally given~by:
\begin{equation}
\hat{r}_{u,i} = \mathbf{w}_u^{\mathsf{T}} ( \mathbf{h}_i + \phi_{\theta}(\mathbf{x}_i)),
\label{eq:cvae}
\end{equation}
where $\mathbf{h}_i$ corresponds to the item embedding obtained in a collaborative filtering framework. However, this approach relies on a shallow user/item interaction model, since the collaborative filtering part remains matrix factorization-based.

More recently, knowledge graph-enhanced recommender systems have been proposed~\citep{Wang2019,Wang2020,Li2020}. These consist in exploiting some latent information that might be shared between items (e.g., the composer of a given song might also have produced other records) in order to yield more explainable content feature extraction, and consequently improving recommendation performance. However, these approaches also consider a shallow user/item interaction model since predictions are made through the dot product of the embeddings.

\subsection{Deep learning-based collaborative filtering}
\label{sec:background_ncf}

In the recent years, leveraging deep learning in collaborative filtering-based recommender systems has attracted a lot of attention~\citep{Zhang2017}. The core idea of such approaches is to replace a shallow matrix factorization model such as presented in Section~\ref{sec:background_wmf} with a DNN. Such a framework allows for learning refined user/item embeddings~\citep{Xue2017} and more complex interaction between these factors~\citep{He2017,Chen2019}. More specifically,~\citep{He2017} proposes to integrate the user and item factors as embeddings in a neural network, which are then jointly learned using a gradient descent (GD) algorithm. First, they propose a model termed generalized matrix factorization (GMF), in which the predicted binarized playcounts are modeled as:
\begin{equation}
    \hat{r}_{u,i} = \sigma( \mathbf{a}^{\mathsf{T}} ( \mathbf{w}_u \odot \mathbf{h}_i ) ),
    \label{eq:gmf}
\end{equation}
where $\sigma$ is a non-linear activation function which replaces the identity function used in matrix factorization~\eqref{eq:wmf_nocontent}, and $\mathbf{a}$ is a weight vector. Alternatively, they consider a concatenation of the user and item factors (instead of their dot product), on top of which a multilayer perceptron (MLP) is applied to learn complex interactions between these factors. They also combine both approaches, yielding a model termed neural matrix factorization. This approach based on learning deep user/item interaction models is shown to outperform the shallow matrix factorization model, which has further been confirmed in~\citep{Chen2019}. However, this so-called neural collaborative filtering (NCF) framework does not use any extra content information, and as such, face the cold-start problem. Several attempts were made to incorporate content information in such a framework~\citep{Lian2017}, but do not leverage the full potential of deep learning. Besides, their potential has been shown for a traditional collaborative filtering task, but not for cold-start recommendation.

We propose to overcome this issue by introducing a unified model in which deep learning is leveraged for both learning deep user/item interactions (as presented in this section) and extracting content features (as described in Section~\ref{sec:background_wmf_content} for cold-start recommendation).

\section{Neural content-aware collaborative filtering}
\label{sec:models}

\subsection{Model overview}
\label{sec:models_overview}

\subsubsection{Data}
\label{sec:models_data}

We consider implicit feedback data in the form of a playcount matrix $\mathbf{Y} \in \mathbb{R}^{U \times I}$. As recalled in Section~\ref{sec:background_wmf}, in order to better account for the over-dispersed nature of such data, it is common to consider the binarized playcount matrix $\mathbf{R}$ rather than the raw listening history. $\mathbf{R}$ indicates whether a user has listened to a song more than a certain amount of times $\tau$ or not~\citep{Liang2015,Liang2016,Tran2019}:
\begin{equation}
\forall (u, i) \text{, } r_{u,i}=
  \left\{
    \begin{aligned}
    & 1 & \text{if } y_{u,i} \geq \tau, \\ 
    & 0 & \text{otherwise.}
    \end{aligned}
  \right.
  \label{eq:binarized_data}
\end{equation}
In order to keep some information about the raw playcounts (which is lost when applying binarization), we also compute a confidence weight defined as:
\begin{equation}
   \forall (u, i) \text{, }  c_{u,i} = 1 + \alpha \log \left( 1 + \frac{y_{u,i}}{\epsilon} \right),
    \label{eq:confidence}
\end{equation}
whose parameters are commonly set at $\alpha=2$ and $\epsilon=10^{-6}$ in the literature~\citep{Hu2008,Liang2015}. Note that alternative schemes exist for defining the confidence, such as using constant values~\citep{Wang2015, Li2017}. We resorted to using~\eqref{eq:confidence} as it yielded slightly better results in our preliminary experiments.

\subsubsection{Generative process}
\label{sec:models_gen}

We model the binarized playcounts as the result of the interaction between the user and item factors. Similarly as in WMF, we consider a Gaussian generative model:
\begin{equation}
    \forall (u, i) \text{, } r_{u,i} \sim \mathcal{N}( \psi_{\gamma}(\mathbf{w}_{u}, \mathbf{h}_{i}) , c_{u,i}^{-1} ),
    \label{eq:wmf_nocontent}
\end{equation}
where $c_{u,i}$ is the confidence defined in~\eqref{eq:confidence} and $\psi_\gamma$ is the interaction model, which might depend on some parameters $\gamma$. The structure of this interaction model will be specifically described in Sections~\ref{sec:models_mf} and~\ref{sec:models_deep}.

As in~\citep{Liang2015}, we consider the following prior on the user factor:
\begin{equation}
    \forall u \in \{1,...,U \} \text{, } \mathbf{w}_{u} \sim \mathcal{N}( 0,  \lambda_W^{-1} \mathbf{I}_K ),
    \label{eq:prior_w}
\end{equation}
where $\lambda_W$ is a regularization hyperparameter. In order to exploit some content information, one needs to consider an additional assumption about the item factor $\mathbf{h}_{i}$. Let us first consider a \emph{relaxed} formulation, where this content information is incorporated in the model in the form of a prior as:
\begin{equation}
    \forall i \in \{1,...,I \} \text{, } \mathbf{h}_{i} \sim \mathcal{N}( \phi_{\theta}(\mathbf{x}_i),  \lambda_H^{-1} \mathbf{I}_K ),
    \label{eq:prior_h_deep}
\end{equation}
where $\phi_\theta$ is a deep content feature extractor with parameters $\theta$ and $\mathbf{x}_i$ is a vector of low-level acoustic features for item $i$. We also consider a \emph{strict} formulation, where the item attribute is directly predicted by the deep content feature extractor:
\begin{equation}
    \forall i \in \{1,...,I \} \text{, } \mathbf{h}_{i} = \phi_{\theta}(\mathbf{x}_i),
    \label{eq:h_strict}
\end{equation}
which corresponds to~\eqref{eq:prior_h_deep} with $\lambda_H \to \infty$.

\subsubsection{Deep content feature extractor}
\label{sec:models_content}

As for the deep content feature extractor $\phi_{\theta}$, we consider an MLP architecture with $P$ layers, that is:
\begin{equation}
    \phi_{\theta}(\mathbf{x}_i) = \phi^P (\phi^{P-1} (... \phi^1(\mathbf{x}_i  ))),
\end{equation}
where $\mathbf{x}_i$ is a set of low-level acoustic features (described in Section~\ref{sec:protocol_features}), and $\phi^p$ denotes the $p$-th layer of $\phi_{\theta}$ (for a clarity purpose, we do not explicitly write that $\phi^p$ depends on some parameters $\theta^p$), such that:
\begin{equation}
     \forall p \in \{1,...,P \} \text{, } \phi^p(\mathbf{z} ) = \sigma^p( \mathbf{A}^p  \mathbf{z} + \mathbf{b}^p),
     \label{eq:deep_content_mlp}
\end{equation}
where $\mathbf{A}^p$, $\mathbf{b}^p$ and $\sigma^p$ respectively denote the weights, biases and activation function of the $p$-th layer. The activation function $\sigma^p$ is chosen as the rectified linear unit (ReLU) function for all layers except for the last one, which uses the identity function. The choice of ReLU over alternative non-linear activation functions (such as the hyperbolic tangent or the sigmoid) is motivated by their consistent performance in the literature, notably their capability to reduce the gradient vanishing and overfitting problems~\citep{Glorot2011}. Based on preliminary validation experiments and on prior work~\citep{Liang2015}, each layer uses $1024$ neurons, except for the last one which outputs the item factor of dimension $K$. We consider a total of $P=3$ layers in our experiments.

The goal of this paper is to assess the potential of combining deep content feature extraction with deep interaction modeling in a unified content-aware collaborative filtering framework, rather than specifically optimizing the networks architectures. As such, we leave the usage of more advanced networks (e.g., convolutional layers potentially acting on raw audio waveforms~\citep{Lee2018}, knowledge graph-based architectures~\citep{Wang2020}) to future work.

\subsubsection{Estimation}
\label{sec:models_estim}

Estimating the whole model in the maximum a posteriori sense results in the following optimization problem for the relaxed variant:
\begin{equation}
 \min_{ \xi_{\text{R}}} \mathcal{L}_{\text{NCACF-R}}(\xi_{\text{R}}) := \sum_{u,i} c_{u,i} ( r_{u,i} - \psi_{\gamma}(\mathbf{w}_{u}, \mathbf{h}_{i}) )^2 + \lambda_W \sum_u || \mathbf{w}_u ||^2 + \lambda_H \sum_i || \mathbf{h}_i - \phi_{\theta}(\mathbf{x}_i) ||^2,
    \label{eq:opt_pb_relaxed}
\end{equation}
where $\xi_{\text{R}} = \{ \theta, \gamma, \mathbf{W}, \mathbf{H}\}$ is the whole set of parameters. The problem for the strict variant writes similarly:
\begin{equation}
 \min_{\xi_{\text{S}}} \mathcal{L}_{\text{NCACF-S}}(\xi_{\text{S}}) :=  \sum_{u,i} c_{u,i} \left( r_{u,i} - \psi_{\gamma}(\mathbf{w}_{u}, \phi_{\theta}(\mathbf{x}_i) ) \right)^2 + \lambda_W \sum_u || \mathbf{w}_u ||^2.
    \label{eq:opt_pb_strict}
\end{equation}
with $\xi_{\text{S}} = \{ \theta, \gamma, \mathbf{W}\}$. Depending on the design choice for the interaction model $\psi_{\gamma}$, we can propose several optimization schemes for problems~\eqref{eq:opt_pb_relaxed} and~\eqref{eq:opt_pb_strict}. In general, we will rely on a gradient descent strategy, but when the interaction model reduces to a shallow dot product, it becomes possible to leverage closed-form updates for estimating $\mathbf{W}$ and $\mathbf{H}$ (see Section~\ref{sec:models_mf_hybrid}).

\subsubsection{Recommendation}
\label{sec:models_recom}

Once the model is trained, cold-start recommendation is performed by computing the predicted binarized playcounts for all user/item pairs through:
\begin{equation}
 \hat{r}_{u,i}= \psi_{\gamma}(\mathbf{w}_{u}, \phi_{\theta}(\mathbf{x}_i)).
    \label{eq:pred_cold}
\end{equation}
The proposed model can also be used for a traditional collaborative filtering task which does not suffer from the cold-start problem. Warm-start recommendations is still performed using~\eqref{eq:pred_cold} for the strict variant. However, for the relaxed variant, it is usually preferred~\citep{Liang2015} to directly use the item embedding extracted from collaborative filtering $\mathbf{h}_i$, and to perform recommendation through:
\begin{equation}
\hat{r}_{u,i} = \psi_{\gamma}(\mathbf{w}_{u}, \mathbf{h}_i),
\label{eq:pred_ncacf_in}
\end{equation}
which we will use in our experiments in the warm-start setting.

\subsection{Linear interaction model}
\label{sec:models_mf}

\begin{figure}[t]
    \centering
    \includegraphics[width=.99\columnwidth]{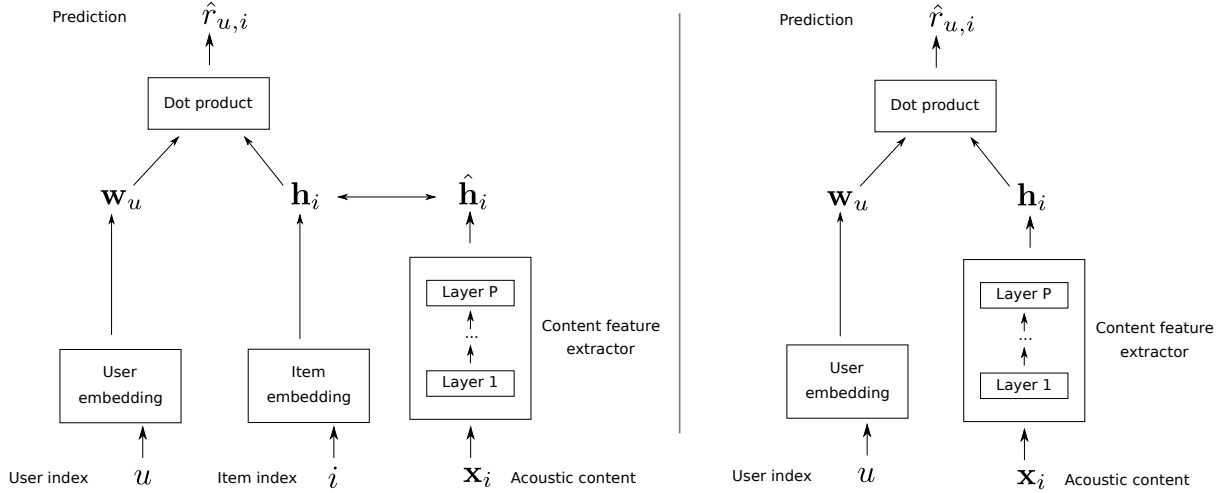}
    \caption{Proposed NCACF models in the specific case where the interaction model reduces to a dot product between the user and item embeddings, in its relaxed (left) and strict (right) variants.}
    \label{fig:ncacf_mf}
\end{figure}

Let us first consider a particular case of the model described above, where the user/item interaction~$\psi_{\gamma}$ reduces to a matrix factorization model:
\begin{equation}
     \psi_{\gamma}(\mathbf{w}_{u}, \mathbf{h}_{i}) = \mathbf{w}_{u}^{\mathsf{T}} \mathbf{h}_{i}.
\end{equation}
The corresponding model is illustrated in Fig.~\ref{fig:ncacf_mf} in its relaxed and strict variants. In such a scenario, the sets of parameters reduce to $\xi_{\text{R}} = \{ \theta, \mathbf{W}, \mathbf{H}\}$ and $\xi_{\text{S}} = \{ \theta, \mathbf{W}\}$, and problems~\eqref{eq:opt_pb_relaxed} and~\eqref{eq:opt_pb_strict} respectively rewrite:
\begin{equation}
 \min_{ \xi_{\text{R}}} \mathcal{L}_{\text{MF-R}}(\xi_{\text{R}}) := \sum_{u,i} c_{u,i} ( r_{u,i} - \mathbf{w}_{u}^{\mathsf{T}} \mathbf{h}_{i} )^2 + \lambda_W \sum_u || \mathbf{w}_u ||^2 + \lambda_H \sum_i || \mathbf{h}_i - \phi_{\theta}(\mathbf{x}_i) ||^2,
    \label{eq:opt_pb_relaxed_mf}
\end{equation}
and
\begin{equation}
 \min_{\xi_{\text{S}}} \mathcal{L}_{\text{MF-S}}(\xi_{\text{S}}) :=  \sum_{u,i} c_{u,i} ( r_{u,i} -  \mathbf{w}_{u}^{\mathsf{T}} \phi_{\theta}(\mathbf{x}_i) )^2 + \lambda_W \sum_u || \mathbf{w}_u ||^2.
    \label{eq:opt_pb_strict_mf}
\end{equation}
To address~\eqref{eq:opt_pb_relaxed_mf} and~\eqref{eq:opt_pb_strict_mf}, we propose two optimization strategies, which we describe hereafter.

\subsubsection{Hybrid algorithm}
\label{sec:models_mf_hybrid}

First, we propose a hybrid approach termed MF-Hybrid which boils down to combining two different algorithms in an alternating minimization framework. Similarly to the WMF model and its content-aware counterpart presented in Section~\ref{sec:background_wmf} and~\ref{sec:background_wmf_content}, this strategy consists in estimating the user and item factors using an ALS scheme, which yields closed-form updates for $\mathbf{W}$ and $\mathbf{H}$ at each iteration. On the other hand, the parameters $\theta$ of the deep content network $\phi$ are estimated using a GD algorithm. This yields the following update scheme for the relaxed version:
\begin{align}
    \forall u \in \{1,...,U \} \text{, } \mathbf{w}_u & \leftarrow (\mathbf{H} \text{diag}(\mathbf{c}_u) \mathbf{H}^{\mathsf{T}} + \lambda_W  \mathbf{I}_K )^{-1} \mathbf{H} \text{diag}(\mathbf{c}_u) \mathbf{r}_u, \label{eq:als_mf_W} \\
    \forall i \in \{1,...,I \} \text{, } \mathbf{h}_i & \leftarrow (\mathbf{W} \text{diag}(\mathbf{c}_i) \mathbf{W}^{\mathsf{T}} + \lambda_H  \mathbf{I}_K )^{-1} (\mathbf{W} \text{diag}(\mathbf{c}_i) \mathbf{r}_i + \lambda_H \phi_{\theta}( \mathbf{x}_i ) ), \label{eq:als_mf_H}\\
    \theta & \leftarrow  \theta - \eta \nabla_{\theta} \left( \sum_i || \mathbf{h}_i - \phi_{\theta}(\mathbf{x}_i)  ||^2 \right),  \label{eq:MSEattributes}
\end{align}
where $\nabla$ denotes the gradient operator, $\eta$ is the learning rate, $\mathbf{r}_u = [r_{u,1},...,r_{u,I}]^{\mathsf{T}}$, $\mathbf{r}_i = [r_{1,i},...,r_{U,i}]^{\mathsf{T}}$, and similarly for $\mathbf{c}_u$ and $\mathbf{c}_i$.\footnote{While $\mathbf{r}_i$ denotes the $i$-th column of $\mathbf{R}$, $\mathbf{r}_u$ denotes its $u$-th row, which might appear as a slight notation abuse. Indeed, using the same notation convention, the $u$-th row should be denoted by $[\mathbf{R}^{\mathsf{T}}]_u^{\mathsf{T}}$. Nonetheless, we decided to keep the notation $\mathbf{r}_u$ for brevity.} Using a similar approach for the strict variant yields:
\begin{align}
    \forall u \in \{1,...,U \} \text{, } \mathbf{w}_u & \leftarrow ( \phi_{\theta}(\mathbf{X}) \text{diag}(\mathbf{c}_i)  \phi_{\theta}(\mathbf{X})^{\mathsf{T}} + \lambda_W  \mathbf{I}_K )^{-1} \phi_{\theta}(\mathbf{X}) \text{diag}(\mathbf{c}_i) \mathbf{r}_u, \label{eq:als_mf_W_strict} \\
    \theta & \leftarrow   \theta - \eta \nabla_{\theta} \left( \sum_{u,i} c_{u,i} ( r_{u,i} - \mathbf{w}_u^{\mathsf{T}} \phi_{\theta}(\mathbf{x}_i) )^2 \right). \label{eq:wpe_strict}
\end{align}
Note that in practice, both~\eqref{eq:MSEattributes} and~\eqref{eq:wpe_strict} might be implemented differently as one might use a stochastic variant of a GD algorithm with momentum. The proposed procedures are termed MF-Hybrid-Relaxed and MF-Hybrid-Strict and summarized in Algorithms~\ref{al:MFhybrid_relaxed} and~\ref{al:MFhybrid_strict}, respectively.
These are iterative schemes in which the first stage of each iteration consists in updating the embeddings using ALS updates, and the second stage consists in updating the network parameters.
Note that the DNN parameters $\theta$ update uses an arbitrary number $N_{\text{gd}}$ of GD steps at each iteration. Indeed, in practice, one does not need to fully minimize the objective with respect to $\theta$ (at $\mathbf{W}$ and $\mathbf{H}$ fixed), but rather seeks for a decrease of the objective function. We will investigate the impact of this parameter in our experiments.

Therefore, these algorithms generalize the related work~\citep{Oord2013}, which consists in first estimating the WMF model in a content-free scenario (that is, using~\eqref{eq:als_mf_W} and~\eqref{eq:als_mf_H} with $\phi_{\theta}=0$) and then learning the mapping $\phi_{\theta}$ through either~\eqref{eq:MSEattributes} or~\eqref{eq:wpe_strict} (the corresponding losses were termed ``mean square error" and ``weighted prediction error" in~\citep{Oord2013}, respectively).

\begin{algorithm}[t]
	\caption{MF-Hybrid-Relaxed}
	\label{al:MFhybrid_relaxed}
			\textbf{Inputs}: Binarized playcounts $\mathbf{R} \in [0, 1]^{U \times I}$ and confidence $\mathbf{C} \in \mathbb{R}_+^{U \times I}$ \\
			Initial user preferences $\mathbf{W} \in \mathbb{R}^{K \times U}$ and item attributes $\mathbf{H} \in \mathbb{R}^{K \times I}$ matrices\\
		    Initial deep content feature network parameters $\theta$ \\
		    Number of iterations $N$, and number of epochs per iteration $N_{\text{gd}}$ \\
			\For{$j=1$ to $N$}{
			 Update $\mathbf{W}$ using~\eqref{eq:als_mf_W} \\
			 Update $\mathbf{H}$ using~\eqref{eq:als_mf_H}\\
			\For{$j'=1$ to $N_{\text{gd}}$}{
			Update $\theta$ using~\eqref{eq:MSEattributes}
			}
			}
			\textbf{Output}: $\mathbf{W}$, $\mathbf{H}$ and $\theta$.
\end{algorithm}

\begin{algorithm}[t]
	\caption{MF-Hybrid-Strict}
	\label{al:MFhybrid_strict}
			\textbf{Inputs}: Binarized playcounts $\mathbf{R} \in [0, 1]^{U \times I}$ and confidence $\mathbf{C} \in \mathbb{R}_+^{U \times I}$ \\
			Initial user preferences matrix $\mathbf{W} \in \mathbb{R}^{K \times U}$\\
		    Initial deep content feature network parameters $\theta$ \\
		    Number of iterations $N$, and number of epochs per iteration $N_{\text{gd}}$ \\
			\For{$j=1$ to $N$}{
            Update $\mathbf{W}$ using~\eqref{eq:als_mf_W_strict}\\
            \For{$j'=1$ to $N_{\text{gd}}$}{
			Update $\theta$ using~\eqref{eq:wpe_strict}
			}
			}

			\textbf{Output}: $\mathbf{W}$ and $\theta$.
\end{algorithm}

\subsubsection{Unified learning}
\label{sec:models_mf_uni}

We now address the optimization problems~\eqref{eq:opt_pb_relaxed_mf} and~\eqref{eq:opt_pb_strict_mf} using a single GD algorithm. We integrate the user and item factors as embedding layers within a DNN, as illustrated in Fig.~\ref{fig:ncacf_mf}. This network consists of a collaborative filtering part and a content feature extractor part. The collaborative filtering part is fed with user and item indices to yield user and item factors through an embedding layer. These are then combined using an interaction model (which here reduces to a dot product) to yield predicted binarized playcounts. The item factor can be regularized using the deep content feature extractor branch (in the relaxed variant) or directly predicted by this branch (in the strict variant).

Even though the resulting models are equivalent to those presented in Section~\ref{sec:models_mf_hybrid}, this now allows for a unified learning approach where only the GD algorithm is used to train all the parameters, by considering as objective functions the losses given in~\eqref{eq:opt_pb_relaxed_mf} and~\eqref{eq:opt_pb_strict_mf}. As a result, the gradient updates are:
\begin{equation}
     \xi_{\text{R}}  \leftarrow   \xi_{\text{R}} - \eta \nabla_{ \xi_{\text{R}} }  \mathcal{L}_{\text{MF-R}}( \xi_{\text{R}}),
\end{equation}
and
\begin{equation}
     \xi_{\text{S}} \leftarrow  \xi_{\text{S}} - \eta \nabla_{\xi_{\text{S}} } \mathcal{L}_{\text{MF-S}}(\xi_{\text{S}}).
\end{equation}
The corresponding procedures will be referred to as MF-Uni-Relaxed and MF-Uni-Strict, respectively.

\subsection{Deep interaction model}
\label{sec:models_deep}

\begin{figure}[t]
    \centering
    \includegraphics[width=.99\columnwidth]{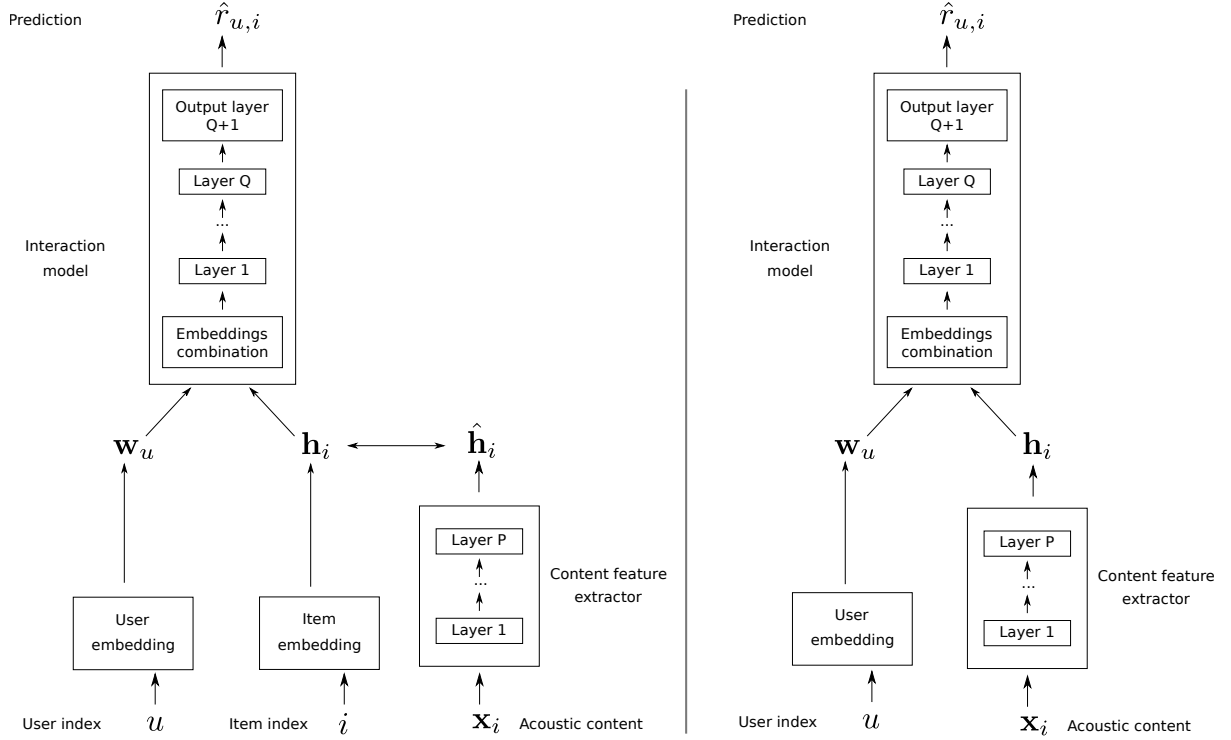}
    \caption{Proposed NCACF model in the general case, where a deep interaction model is used in addition to the deep content feature extractor, in its relaxed (left) and strict (right) variants.}
    \label{fig:ncacf_deep}
\end{figure}

We now propose a more general version of the network where we leverage a deep interaction model in order to learn more refined interactions between embedding vectors~\citep{He2017,Xue2019}. This approach is illustrated in Fig.~\ref{fig:ncacf_deep}. First, the two embeddings are combined into a single vector. Drawing on prior work~\citep{Liu2015,He2017,Xue2019}, we propose two embedding combinations, based on their multiplication or concatenation:
\begin{equation}
\forall (u, i) \text{, } \mathbf{v}_{u,i} =
  \left\{
    \begin{aligned}
    & \mathbf{w}_{u} \odot \mathbf{h}_{i} & \text{ (multiplication)}, \\ 
    & \begin{bmatrix}
\mathbf{w}_{u} \\
\mathbf{h}_{i}
\end{bmatrix} & \text{ (concatenation).}
    \end{aligned}
  \right.
  \label{eq:vec_combination}
\end{equation}
The resulting combined vector $\mathbf{v}_{u,i}$ is of length $K'$, which is equal to $K$ or $2K$ in the case of a multiplication or concatenation, respectively. Previous studies such as~\citep{Chen2019} have shown that using a concatenation of the user and item factors overall outperforms using a multiplication, regardless of the size and number of deep layers of the subsequent network. However, since our framework is different (we consider the cold-start scenario and we use an additional deep content extractor), we will evaluate both combination methods experimentally. This combined vector is then fed as input to an MLP with $Q$ hidden layers to output the predicted binarized playcounts, which is defined similarly as the deep content feature extractor~\eqref{eq:deep_content_mlp}:
\begin{equation}
    \hat{r}_{u,i} = \psi^{Q+1}(\psi^{Q} (... \psi^{1}(\mathbf{v}_{u,i} ))),
\end{equation}
where $\psi^{q}$ denotes the $q$-th layer of the network $\psi_\gamma$, such that:
\begin{equation}
     \forall q \in \{1,...,Q+1\} \text{, } \psi^q(\mathbf{z} ) = \sigma^q(\mathbf{A}^q \mathbf{z} + \mathbf{b}^q),
\end{equation}
where $\mathbf{A}^q$, $\mathbf{b}^q$ and $\sigma^q$ are the weights, biases and activation function of the $q$-th layer, respectively (we use the same notations as for the the deep content feature extractor $\phi_\theta$ for clarity, but the corresponding parameters are different). The last layer (i.e., the output layer) has a particular structure, since it does not use biases (that is, ${\mathbf{b}^{Q+1} = 0}$) and uses a single neuron in order to output the predicted binarized playcount of dimension $1$. As activation functions, we choose ReLU for the $Q$ hidden layers and the sigmoid function for the output layer, which is chosen to enforce the predicted binarized playcounts to range between $0$ and $1$~\citep{He2017,Chen2019}.

In order to choose an appropriate number of neurons for each hidden layer, we resort to using the so-called ``tower" structure, which has shown appropriate in deep collaborative filtering models~\citep{He2016a, He2017,Chen2019}. Specifically, this structure consists in halving the size of each successive higher layers, as this allows for learning more abstract representations in higher layers of the network~\citep{He2017}. In practice, the $q$-th hidden layer therefore uses $K' / 2^{q-1}$ neurons.

These models will be referred to as NCACF-Relaxed and NCACF-Strict, and are estimated by addressing the optimization problems~\eqref{eq:opt_pb_relaxed} and~\eqref{eq:opt_pb_strict} using a GD algorithm, which yields the following updates:
\begin{equation}
    \xi_{\text{R}}  \leftarrow  \xi_{\text{R}} - \eta \nabla_{\xi_{\text{R}} }  \mathcal{L}_{\text{NCACF-R}}(\xi_{\text{R}}),
\end{equation}
and
\begin{equation}
    \xi_{\text{S}}  \leftarrow  \xi_{\text{S}} - \eta \nabla_{\xi_{\text{S}} } \mathcal{L}_{\text{NCACF-S}}(\xi_{\text{S}}),
\end{equation}
where $\mathcal{L}_{\text{NCACF-R}}$ and $\mathcal{L}_{\text{NCACF-S}}$ are the losses defined in~\eqref{eq:opt_pb_relaxed} and~\eqref{eq:opt_pb_strict}, respectively. In particular, when the interaction model is based on a multiplication of the embeddings and uses a single output layer with a fixed weight vector $\mathbf{a} = \mathbf{1}$ and no activation function (i.e., the sigmoid is replaced with the identity), this model becomes equivalent to the MF-Uni model presented in Section~\ref{sec:models_mf_uni}.

\subsection{Training considerations}
\label{sec:model_imp}

\subsubsection{Loss functions}

The proposed NCACF model and its variants are based on a Gaussian generative process for the playcounts, from which a weighted prediction error naturally arises as loss function (\textit{cf}.~\eqref{eq:opt_pb_relaxed} and~\eqref{eq:opt_pb_strict}). However, this loss is known not to be the most appropriate choice when handling implicit feedback data such as playcounts~\citep{Salakhutdinov2007a}. Consequently, other losses such as the log loss have been preferred in such tasks~\citep{He2017,Xue2017,Chen2019}. Nonetheless, we decided to use the weighted prediction error in this work since the underlying Gaussian generative model allows to easily incorporate prior information, which has motivated previous work on content-aware collaborative filtering such as~\citep{Wang2015,Wang2016rnn,Huan2017} to adopt this framework. Besides, alternative losses might improve the overall performance of all proposed methods, but it would make the comparison unfair with baselines such as~\citep{Oord2013}.

Therefore, we leave the usage of alternative losses to future work. Note that this would be consistent with the design of alternative generative models, for instance by directly modeling the raw interaction data instead of the binarized playcounts. In particular, refined statistical models based on the Poisson or compound Poisson distributions~\citep{Gopalan2014,Gopalan2015,Basbug2016,Gouvert2019} allow for robust modeling of such over-dispersed implicit feedbacks, and might yield more appropriate losses for this task.

\subsubsection{Sampling} 

Except for MF-Hybrid-Relaxed, learning the proposed models involves to perform GD updates for minimizing a weighted prediction error on the form of:
\begin{equation}
 \sum_{u=1}^U \sum_{i=1}^I c_{u,i} ( r_{u,i} - \hat{r}_{u,i} )^2.
    \label{eq:wpe_general}
\end{equation}
In practice, this problem is split over batches of data $\mathcal{B} \subset \{1,..,U\} \times \{1,..,I\} $, each of which being processed at each iteration of the GD algorithm.  The loss then rewrites for each batch:
\begin{equation}
 \sum_{(u,i)\in \mathcal{B}} c_{u,i} ( r_{u,i} - \hat{r}_{u,i} )^2.
    \label{eq:wpe_batch}
\end{equation}
Processing all the available data (both non-zero and zero playcounts) results in a relatively high computational burden, which might become cumbersome for very large datasets. To alleviate this issue, and to account for the sparsity of the dataset, it is common to resort to a negative sampling strategy~\citep{Hsieh2017,Lee2018,Chen2019}, where only part of the null playcounts is considered, that is:
\begin{equation}
 \sum_{(u,i)\in \mathcal{B}^+} \left[ c_{u,i} ( r_{u,i} - \hat{r}_{u,i} )^2 +   \sum_{j \in \mathcal{B}_u^-} c_{u,j} ( r_{u,j} - \hat{r}_{u,j} )^2 \right]
    \label{eq:wpe_posneg}
\end{equation}
where $\mathcal{B}^+$ is a batch of positive samples (that is, a set of user/item pairs with a non-zero playcount) and $\mathcal{B}_u^- = \{ j=1, \ldots ,N^- | r_{u,j} = 0 \}$ is a set of $N^-$ negative samples for user $u$. We tested this strategy in preliminary experiments, but we obtained a relatively poor performance for all methods. Consequently, we rather consider a more traditional sampling strategy, where we considered batches of items:\footnote{Note that an alternative common choice consists in considering batches of users instead of items. However, in a content-aware framework, considering batches of items is more straightforward to train both the collaborative filtering part and the content extractor part, since the latter operates on items (and not on users).}
\begin{equation}
 \sum_{i\in \mathcal{B}_{\mathsf{i}} } \sum_{u=1}^U c_{u,i} ( r_{u,i} - \hat{r}_{u,i} )^2
    \label{eq:wpe_batchitem}
\end{equation}
where $\mathcal{B}_{\mathsf{i}} \subset \{1,..,I\}$.
Even though this approach imposes to use a smaller batch size, since playcounts for all users are considered for each sample $i \in \mathcal{B}_{\mathsf{i}}$, it yields a significantly better performance for all tested methods. Consequently, we present the results obtained using this sampling strategy.

\subsection{Relation to other works}
\label{sec:model_relation}

The proposed framework encompasses and generalizes several collaborative filtering-based models from the literature. We summarize these in Table~\ref{tab:models_overiew} along with their main characteristics, that is:

\begin{itemize}
    \item whether the model is trained using a two-stage or a joint training approach;
    \item whether the training algorithm hybridizes two forms of updates or a single gradient descent;
    \item whether the model uses a shallow or a deep interaction model;
    \item whether it is content-aware, thus suitable for cold-start recommendation, or not.
\end{itemize}

The NCACF-Relaxed model is notably an extension of the NCF approach~\citep{He2017} with an additional branch to account for some content information, which makes it suitable for cold-start recommendation. In particular, when the interaction model is based on a multiplication of the embeddings and uses a single layer, NCACF reduces to the variant termed generalized matrix factorization (\textit{cf}.~\eqref{eq:gmf}).

When the interaction model in NCACF reduces to a dot product, the corresponding models (MF-Hybrid and MF-Uni) also encompass some previous proposals. In particular, MF-Hybrid-Relaxed is similar to~\citep{Wang2015, Li2017}, as these works use a one-stage approach for learning the model which is based on a hybrid algorithm (using both ALS and GD). Note however that our proposal is slightly more flexible since it allows for using an arbitrary number of GD epochs in between ALS updates (\textit{cf}. Algorithm~\ref{al:MFhybrid_relaxed}). If we further simplify the training procedure of MF-Hybrid by only conducting ALS updates (in a content-free framework) and then learning the deep network $\phi_{\theta}$, then it reduces to the deep content-based (DCB) recommendation technique~\citep{Oord2013} presented in Section~\ref{sec:background_wmf_content}.

On the other hand, MF-Uni-Strict is somewhat equivalent to the deep content-user embedding (DCUE) model introduced in~\citep{Lee2018}. The model~\citep{Zheng2017} is also similar to MF-Uni-Strict when the interaction model is replaced by a factorization machine. These approaches are however not tested in a specific cold-start setting, even though they are suitable for this task, but rather consider a traditional collaborative filtering application. \cite{Lee2018} actually observe that DCUE yields worse results than a simple WMF-based model. This observation is reminiscent of prior works such as~\citep{Liang2015} where it is observed that warm-start recommendation usually does not benefit from adding extra content information. Therefore, we will test this approach in a cold-start scenario, which appears more appropriate for highlighting the potential of this content-aware model.

We also propose MF-Uni-Relaxed, which is a more flexible approach than its strict counterpart. Besides, still according to~\citep{Liang2015}, it is expected to outperform the strict variant~\citep{Lee2018} in a warm-start collaborative filtering. MF-Uni-Relaxed shares some similarity with the content-boosted collaborative filtering network (CCCFNet) presented in~\citep{Lian2017}. It uses a shallow content extractor and exploits two item embeddings (a collaborative filtering one and another one extracted from side-content). Like DCUE, CCFNet has only been evaluated in a warm-start recommendation scenario.

Overall, our unifying framework not only generalizes several methods from the literature, but also allows for making them suitable for cold-start recommendation, for which their potential was partly still left to explore.

\begin{table}[t]
    \centering
    \caption{Overview of the proposed models, related approaches in the literature, and main characteristics.}
    \label{tab:models_overiew}
    \begin{tabular}{lccccc} 
\hline\noalign{\smallskip}
    Model & Related work  & Joint learning & GD-only & Deep interaction & Content-aware \\
\noalign{\smallskip}\hline\noalign{\smallskip}
    NCF &  \citep{He2017} & - & \cmark & \cmark & \xmark \\
\noalign{\smallskip}\hline\noalign{\smallskip}
    DCB   &   &  & &  &  \\
    \multicolumn{1}{r}{Relaxed} & \citep{Oord2013} & \xmark & \xmark & \xmark & \cmark \\
    \multicolumn{1}{r}{Strict}  & \citep{Oord2013} & \xmark & \xmark & \xmark & \cmark \\
    MF-Hybrid   &   &  & &  &  \\
    \multicolumn{1}{r}{Relaxed} &  \citep{Wang2015} & \cmark & \xmark & \xmark & \cmark \\  
    \multicolumn{1}{r}{Strict} & -  & \cmark & \xmark & \xmark & \cmark \\  
    MF-Uni    &   &  & &  &  \\
    \multicolumn{1}{r}{Relaxed}    & \citep{Lian2017}  & \cmark & \cmark & \xmark & \cmark \\
    \multicolumn{1}{r}{Strict}   &  \citep{Lee2018}   & \cmark & \cmark & \xmark & \cmark \\  
    NCACF     &   &  & &  &  \\
    \multicolumn{1}{r}{Relaxed}  &  -    & \cmark & \cmark & \cmark & \cmark \\ 
    \multicolumn{1}{r}{Strict}  & -     & \cmark & \cmark & \cmark & \cmark \\ 
\noalign{\smallskip}\hline
    \end{tabular}
\end{table}

\section{Experimental protocol}
\label{sec:protocol}

In this section, we present the experimental protocol to assess the potential of the proposed model for music recommendation. Even though our framework is primarily intended to perform cold-start recommendation, we also consider a more traditional warm-start scenario, in order to show the potential of exploiting side content information in such a case. In the spirit or reproducible research, we provide the code related to these experiments.\footnote{\url{https://github.com/magronp/ncacf}} Our method is implemented using
the PyTorch framework,\footnote{\url{https://pytorch.org/}} and all the computations have been performed using an NVIDIA Tesla T4 GPU with 15 GB RAM.

\subsection{Research questions}
\label{sec:protocol_questions}

Our experiments aim at answering the following research questions:
\begin{enumerate}
    \item Is a joint training approach preferable over a two-stage training for content-aware models?
    \item How does a unified training algorithm perform compared to a method that hybridizes two forms of updates?
    \item What is the impact of the interaction model (shallow vs. deep) onto performance for both warm- and cold-start recommendation?
    \item Which training loss, or equivalently, which model variant (relaxed or strict) performs the best?
    \item Does a content extractor improves performance even in the warm-start setting?
    \item How do content-aware baselines perform for cold-start recommendation (generally, these have been tested only for warm-start recommendation)?
\end{enumerate}
\noindent Questions 1, 2, and 3, will be addressed in Sections~\ref{sec:results_stages},~\ref{sec:results_hybvsuni}, and~\ref{sec:results_ncacf}, respectively, while Question 4 will be addressed throughout these sections. Questions 5 and 6 will be treated in Section~\ref{sec:results_overall}.

\subsection{Playcount data}
\label{sec:protocol_playcount}

As implicit feedback data, we use the Taste Profile dataset\footnote{\url{http://millionsongdataset.com/tasteprofile/}} which is part of the Million Song Dataset~\citep{BertinMahieux2011}. It provides listening counts of $1$ million users and $380,000$ songs. After removing duplicates, we kept the songs whose acoustic content features were available (see Section~\ref{sec:protocol_features}). The playcount data is binarized by retaining values of seven or higher as implicit feedback (that is, $\tau=7$ in~\eqref{eq:binarized_data}), since lower values yield feedback that are commonly considered as non-informative~\citep{Tran2019}. In accordance with~\eqref{eq:binarized_data}, other values are set at $0$. As in~\citep{Liang2015,Tran2019}, in order to keep the computational burden low, we retain the top songs and users (sorted by playcounts) and we remove inactive users and items: that is, we only keep users who listened to at least 20 songs, and songs which have been listened to by at least 50 users. The resulting dataset has a density of about $0.2$ $\%$, which is a relatively high level of sparsity when compared to other datasets on which similar methods are tested~\citep{Chen2019}.

To train and evaluate the models, we resort to a $10$-fold cross-validation strategy~\citep{Flexer2006}. In the cold-start setting, we hold 20~\% of the songs out as validation subset. We then randomly partition the remaining songs into $10$ equally sized subsets, thus 90~\% and 10~\% of these songs serve for training and testing, respectively.
The playcounts corresponding to the training subset are used for learning the various models. The held-out validation subset is used for tuning the hyperparameters (see Section~\ref{sec:protocol_details}), and the remaining test set is used for comparing the models. Evaluation on the validation and test sets corresponds to the cold-start scenario, since the models have not been trained on the playcounts corresponding to the songs they contain. In order to reduce the computational burden, hyperparameters are only tuned on the first training/testing split. The corresponding optimal values are then used for all $9$ remaining splits when testing.

For warm-start recommendation, we use a similar strategy, except we split the non-zero playcounts instead of the songs: this ensures that all songs in the validation and test sets also have some non-zero interaction in the training set (otherwise, warm-start recommendation would not be possible). The resulting datasets' properties are detailed in Table~\ref{tab:dataset_stat}.

\begin{table}[t]
    \centering
    \caption{Dataset statistics for the first split (interactions denote the number of non-zero playcounts).}
    \label{tab:dataset_stat}
    \begin{tabular}{lc|ccc|ccc} 
\hline\noalign{\smallskip}
    &    & \multicolumn{3}{c}{Warm-start}  & \multicolumn{3}{c}{Cold-start} \\
      &  Total  & Training &  Validation & Test &  Training &  Validation & Test \\
\noalign{\smallskip}\hline\noalign{\smallskip}
      Users         & $17,572$  & $17,572$  & $17,572$  & $17,572$ & $17,572$  & $17,572$  & $17,572$ \\  
      Songs         & $13,532$  & $13,532$  & $13,532$  & $13,532$ & $9,743$   & $2,706$   & $1,083$  \\  
      Interactions  & $509,177$ & $366,607$ & $101,835$ & $40,735$ & $355,891$ & $105,897$ & $47,389$ \\  
\noalign{\smallskip}\hline
    \end{tabular}
\end{table}

\subsection{Acoustic content features}
\label{sec:protocol_features}

Several previous works~\citep{Oord2013,Lee2018} rely on extracting acoustic features from the raw audio waveforms directly. Due to copyright restriction, the audio files corresponding to the songs in the Million Song Dataset cannot be distributed. Therefore, these prior works have relied on the 7digital platform,\footnote{\url{https://us.7digital.com/}} which allows to download short audio samples for small evaluations and prototyping. However, these audio snippets are not easily obtained due to access restrictions of the 7digital-API. Besides, these are limited to 30s-long snippets, which makes it difficult to extract features that encompass the full temporal dynamic of the songs.\footnote{For these reasons, we did not consider alternative datasets such as KKBox (\url{https://www.kaggle.com/c/kkbox-music-recommendation-challenge/data}), for which only the listening history is directly available.} An alternative option consists in using the pre-extracted features provided with the Million Song Dataset and computed using the Echo Nest platform. While a subset of these features is easily available for prototyping, the full dataset is however difficult to access, as it is stored on an Amazon snapshot which access requires a non-free account. Finally, we could not access the Echo Nest API to recompute these features.\footnote{The Echo Nest developer API (\url{http://developer.echonest.com/}) was not accessible at the time of conducting this research.}

In the spirit of reproducible research, we aimed at using easily accessible acoustic features. Therefore, we used the statistical spectrum descriptors (SSDs)~\citep{Lidy2005}, since these features have shown good performance in several music information retrieval tasks such as genre recognition~\citep{Schindler2012,Schindler2012a}, and are freely available online.\footnote{\url{http://www.ifs.tuwien.ac.at/mir/msd/download.html}} The SSDs are sets of statistical moments extracted from the sonogram of each song, a mid-level time-frequency representation that reflects the human loudness sensation. To compute the SSDs, \cite{Schindler2012} first obtained the raw audio waveform of each song using the 7digital platform, from which they obtained a time-frequency representation by applying a short-time Fourier transform. The resulting frequency channels are then grouped into $24$ psycho-acoustically motivated critical bands, accounting for several masking effects. Finally, seven statistical moments (mean, median, variance, skewness, kurtosis, min, and max) are computed in each critical band to account for the temporal dynamic of each song. This results in a set of low-level acoustic features for each song $\mathbf{x}_i \in \mathbb{R}^L$ with $L=168$, which we scale to have $0$ mean and unit variance.

Note that alternative pre-calculated low-level acoustic features are available, such as MFCC-based statistical moments computed with the Essentia toolkit~\citep{Bogdanov2013}.\footnote{\url{https://zenodo.org/record/3258042##.YpYacTnP1so}} However, these are computed from 30s-long excerpts, and stacked into a feature vector of length $L=40$, while SSDs are vectors of length $L=168$ computed from the whole songs. Therefore, we resort to using the SSDs, since they capture more comprehensively the acoustic content of the songs.

\subsection{Training details}
\label{sec:protocol_details}

Following a previous music recommendation work using the same dataset~\citep{Tran2019}, we set the user and item embeddings dimension at $K=128$, whether these are computed using ALS or integrated within the network. The user/item embeddings are initialized (in all models) with random values drawn from a centered normal distribution with a standard deviation of $10^{-2}$. The deep content feature extractor and deep interaction model parameters $\theta$ and $\gamma$ are initialized using the default Le Cun's initialization scheme~\citep{LeCun2012}, except for the output layer of the interaction model in NCACF (and the NCF baseline), whose weights are all initialized with ones. GD is performed using the Adam algorithm~\citep{Kingma2015} with a learning rate of $10^{-4}$ and a batch size of $128$. The hyperparameters $\lambda_W$ and $\lambda_H$ are tuned on the validation set and chosen to maximize the NDCG metric (see Section~\ref{sec:protocol_ndcg}). The number of epochs is determined on the validation set, with a maximum value of $150$. In order to save some computational time, NCACF uses the same hyperparameters as MF-Uni and is pretrained using no deep interaction model nor activation function (which corresponds to MF-Uni when the interaction model is a multiplication of the embeddings) for a maximum number of $100$ epochs. The subsequent NCACF models are then trained for a maximum of $100$ more epochs.

\subsection{Evaluation metric}
\label{sec:protocol_ndcg}

We use the NDCG~\citep{Wang2013} as a measure of the overall quality of the recommendation. For each user $u$, we compute a ranked list of items in the evaluation (i.e., validation or test) set based on the predicted preferences $\hat{\mathbf{r}}_u$. We then compute the relevance of this list with respect to the ground truth preferences, that is, the observed user/item interaction in the evaluation set: $\text{rel}_{u,i} = 1$ if the item $i$ is in the listening history of user $u$ (in the evaluation set) and $0$ otherwise. In order to favor recommendations that place the evaluation items high in the list, we apply a discounted weight to the relevance, which yields the discounted cumulative gain (DCG):
\begin{equation}
    \text{DCG}_u = \sum_{i=1}^{I'} \frac{\text{rel}_{u,i}}{ \log_2(i+1) },
    \label{eq:dcg}
\end{equation}
where $I'$ denotes the length of the list of ranked items. To obtain a metric that accounts for all items in the evaluation set, $I'$ should be equal to the number of songs in this set. However, this considerably increases the computational load to evaluate the methods. Consequently, a common approach consists in considering the truncated list of top-$I'$ items instead~\citep{Liang2016}. Following similar work conducted on this dataset~\citep{Tran2019}, we used $I'=50$ in our experiments, which is reasonable for a music recommendation application, and significantly reduces the computational burden. 
The normalized version of the DCG is then obtained as follows:
\begin{equation}
    \text{NDCG}_u = \frac{\text{DCG}_u}{ \text{IDCG}_u },
\end{equation}
where IDCG is the ideal DCG, which corresponds to the DCG of a perfectly ranked list. Finally, these scores are averaged over users to yield an overall recommendation performance. The resulting NDCG ranges from $0$ to $1$ (higher is better), and will be expressed in $\%$ in our experiments for readability.

\subsection{Baselines}
\label{sec:protocol_baselines}

We compare our proposed NCACF model with the following methods:
\begin{itemize}
    \item \textbf{WMF}~\citep{Hu2008}. This basic matrix factorization-based recommendation method does not leverage any side-content information. It is described in Section~\ref{sec:background_wmf}.
    \item \textbf{DCB}~\citep{Oord2013}. This two-stage deep content-based (DCB) recommendation technique uses a linear interaction model, and it described in Section~\ref{sec:background_wmf_content}.
    \item \textbf{CDL}~\citep{Wang2015}. The collaborative deep learning framework leverages side-content information and relies on a linear interaction model. Unlike DCB, the whole model is trained jointly in a single stage by combining ALS updates and GD iterations, as described in Section~\ref{sec:background_wmf_content}.
    \item \textbf{DCUE}~\citep{Lee2018}. This model is similar to CDL, but it is trained using a unified GD-based algorithm.
    \item \textbf{CCCFNet}~\citep{Lian2017}. This model is similar to DCUE and is also trained using a unified GD algorithm, but it uses two different item embeddings (extracted from the collaborative filtering and side-content, respectively), thus sharing more similarity with our MF-Uni-Relaxed.
    \item \textbf{NCF}~\citep{He2017}. This method exploits a deep interaction model and reaches state-of-the-art warm-start recommendation performance. It is described in Section~\ref{sec:background_ncf}.
\end{itemize}
Note that even though several baseline methods use alternative deep architectures for content extraction (e.g., convolutional layers for DCUE or auto-encoders for CDL), we implement them using the same MLP described in Section~\ref{sec:models_overview} for a fair comparison. The link between these methods and our implementation as variants of the NCACF model is described in Section~\ref{sec:model_relation}.

\section{Results and discussion}
\label{sec:results}

\subsection{Overall recommendation results}
\label{sec:results_overall}

Let us examine the recommendation performance of the baselines and the proposed NCACF model for a warm- and cold-start recommendation tasks. The results presented in Table~\ref{tab:overall_comp} correspond to the best performing method in each category for a fair comparison.

Firstly, by comparing WMF and content-aware approaches using a shallow interaction model, we observe that leveraging side content information does not significantly improves performance in the warm-start setting (except for CCCFNet, for which a slight performance improvement can be noticed), but mostly allow to address the cold-start problem. This is in line with several previous works such as~\citep{Liang2015, Magron2021avd} that suggested that accounting for extra content information was not beneficial in this setting. The slightly different behavior of CCCFNet might be explained by the fact that in this strict approach, acoustic features are used to directly predict the item embedding, thus it exploits the available content information in a way that is more optimal for the recommendation task at hand. Nonetheless, a different trend is observed when comparing NCF and NCACF (whose deep interaction models are similar) in the warm-start setting, since the latter significantly outperforms its content-free counterpart. This outlines the potential of content-aware approaches even in traditional collaborative filtering settings, provided a deep content feature extractor is used.

\begin{table}[t]
    \centering
    \caption{Overall performance of the compared methods on the test set (mean NDCG in \% $\pm$ standard deviation over the 10 folds).}
    \label{tab:overall_comp}
    \begin{tabular}{lcc} 
\hline\noalign{\smallskip}
    & Warm-start & Cold-start \\
\noalign{\smallskip}\hline\noalign{\smallskip}
    WMF         & $31.5 \pm 0.2$ &  $-$  \\
    DCB         & $30.0 \pm 1.8$ & $23.7 \pm 1.7$  \\
    CDL         & $31.3 \pm 0.2$ & $25.3 \pm 0.9$ \\
    DCUE        & $27.3 \pm 0.2$ & $24.7 \pm 1.0$  \\
    CCCFNet     & $31.7 \pm 0.1$ & $23.9 \pm 1.2$  \\
    NCF         & $41.1 \pm 0.4$ &  $-$ \\
    NCACF       & $\textbf{43.6} \pm 0.5$  &  $\textbf{26.9} \pm 4.0$ \\
\noalign{\smallskip}\hline
    \end{tabular}
\end{table}

Let us now compare the various content-aware approaches using a shallow interaction model. DCB is outperformed by the other approaches in all configurations, except for DCUE in the warm-start setting. Overall, this demonstrates the advantage of resorting to a joint learning strategy rather than a two-stage approach. Interestingly, while CCFNet yields the best results in the warm-start setting among these approaches, this is not the case in the cold-start setting, where it is outperformed by CDL. The main difference between these methods being the training algorithm (GD only vs. ALS+GD), leveraging closed-form updates should be preferred to a unified algorithm in the cold-start setting (when at least part of the model is tractable). Besides, note that these two methods (CDL and CCFNet) are relaxed approaches, since the item embedding is regularized by the content network (but not directly predicted by it). This emphasizes the benefit of using such a flexible approach, as it will be confirmed in the following experiments. Finally, we remark that DCUE yields poor results in the warm-setting, where it is outperformed by a basic WMF, a conclusion that is reminiscent of the original DCUE paper~\citep{Lee2018}. However, the results in Table~\ref{tab:overall_comp} also reveal that DCUE outperforms DCB in the cold-start setting: this finding outlines that the true potential of such an approach lies within the cold-start setting, which complements the experiments conducted in~\citep{Lee2018} where only warm-start recommendation was considered.

A comparison between NCF and the other baselines in the warm-start setting reveals the potential of using a deep interaction model. Indeed, NCF yields a large $8.1$ \% NDCG improvement over the best performing approach using a shallow interaction model (CCFNet), which confirms the findings of previous studies~\citep{He2017, Chen2019}. As mentioned above, exploiting content information (i.e., using NCACF) allows to boost the performance even further in this setting. More interestingly, this approach allows to alleviate the cold-start problem faced by NCF: NCACF yields the best cold-start performance, outperforming the previous best performing approach (CDL) by $1.6$ \% in NDCG. This highlights the potential of using a deep user/item interaction model over a conventional matrix factorization-based technique, where both the interaction model and the content extractor are jointly trained for cold-start music recommendation.

\subsection{Learning strategies for a linear interaction model}
\label{sec:results_stages}

The goal of this experiment is to assess the potential of a joint training strategy over a two-stage approach for the variants based on a linear interaction model. To that end, we compare MF-Hybrid and DCB, which correspond to these two strategies. 

\begin{figure}[p]
\centering
    \begin{subfigure}{.9\linewidth}
    \hspace{2.1em} \includegraphics[width=.9\columnwidth]{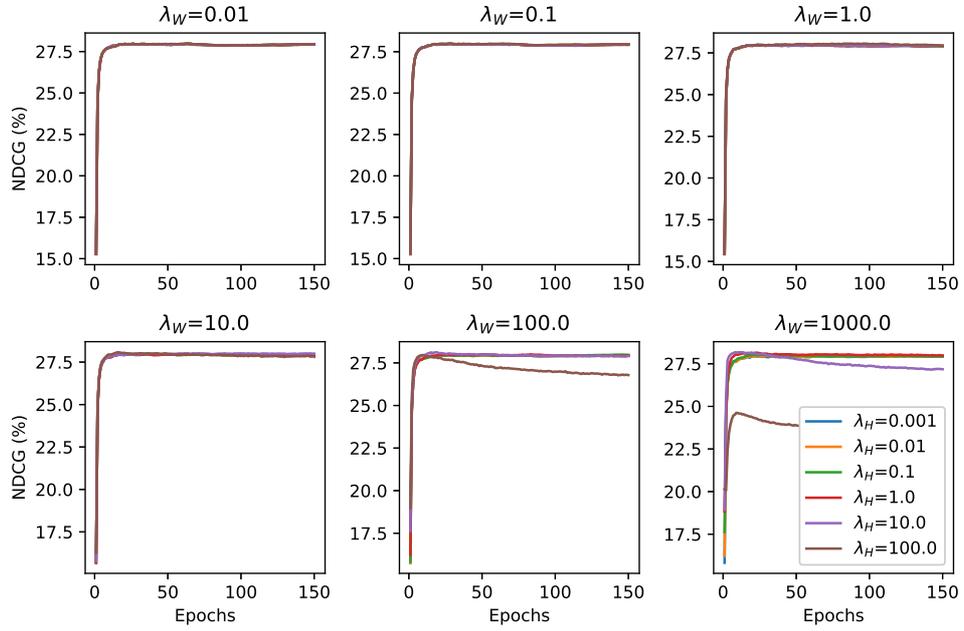}
    \vspace{-0.5em}
    \subcaption{Relaxed}
    \label{fig:mf_hybrid_relaxed_in}
    \end{subfigure}
    \newline
    \begin{subfigure}{.9\linewidth}
    \centering
    \vspace{1em}
    \includegraphics[width=.9\columnwidth]{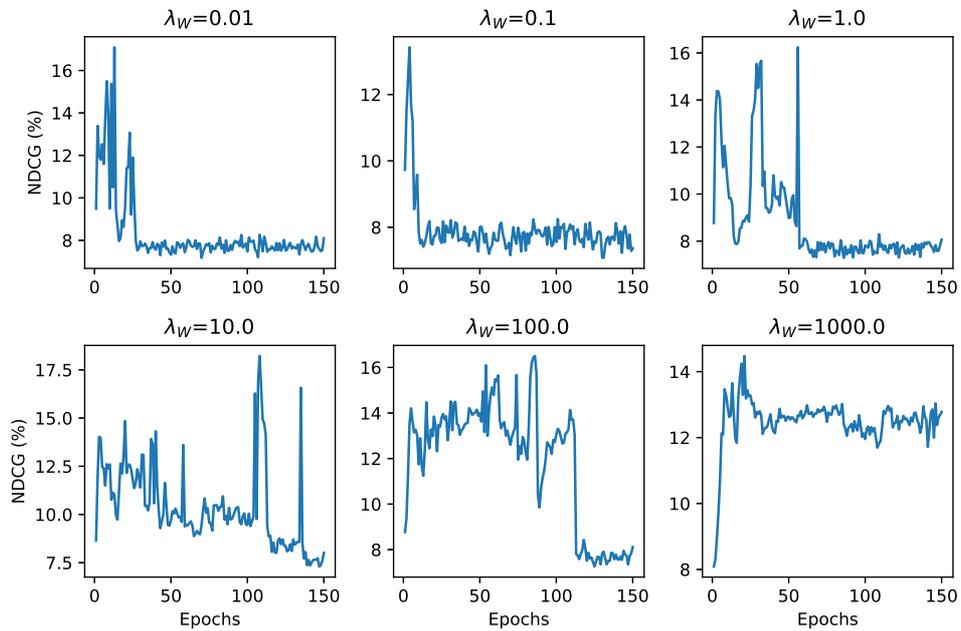}
    \vspace{-0.5em}
    \subcaption{Strict}
    \label{fig:mf_hybrid_strict_in}
    \end{subfigure}
    \caption{NDCG on the validation set for the MF-Hybrid algorithms (relaxed variant on the top, strict variant on the bottom) in the warm-start setting, for several values of the hyperparameters.}
    \label{fig:mf_hybrid_in}
\end{figure}

\begin{figure}[p]
\centering
    \begin{subfigure}{.9\linewidth}
    \hspace{2.1em} \includegraphics[width=.9\columnwidth]{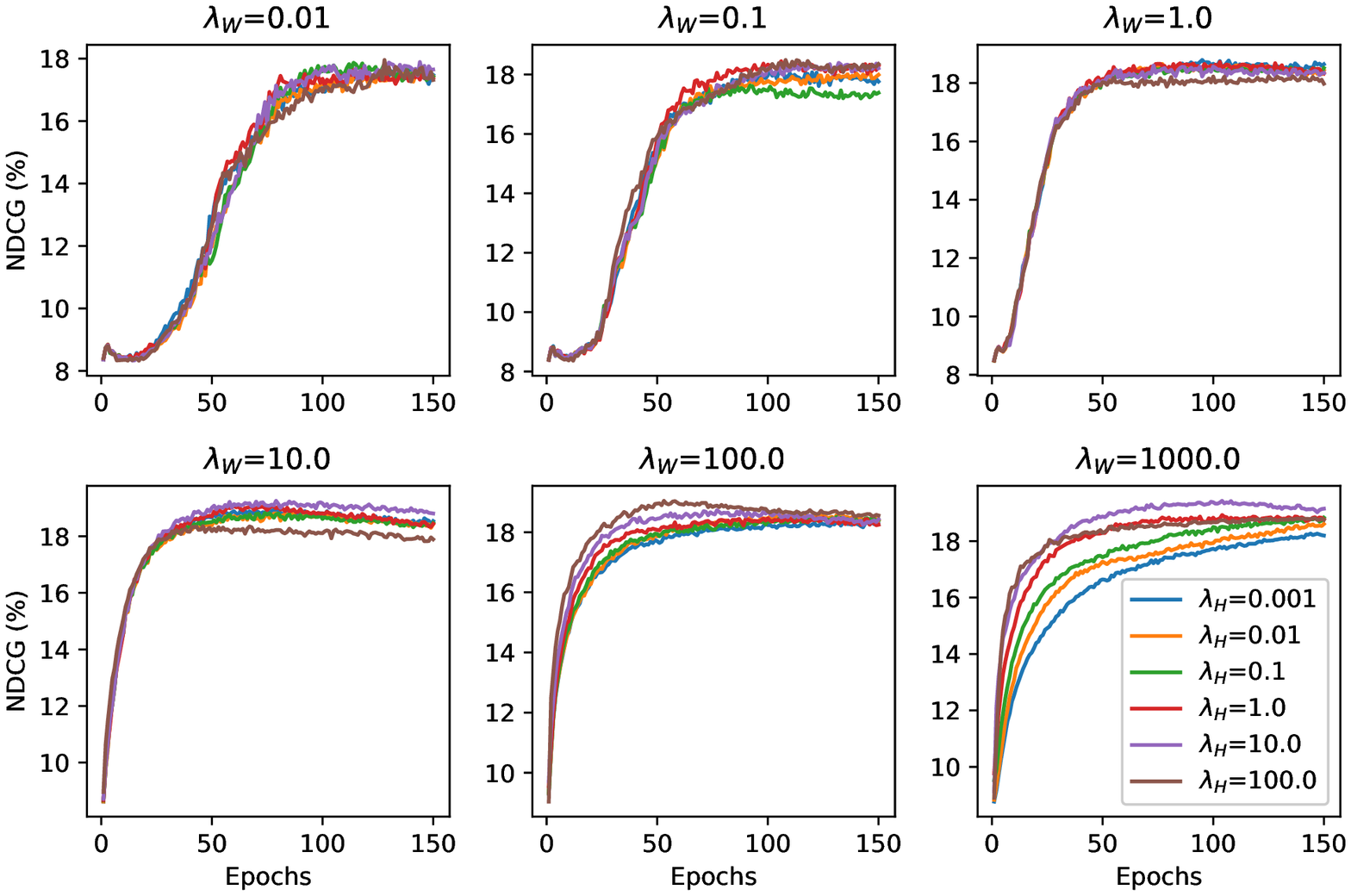}
    \vspace{-0.5em}
    \subcaption{Relaxed}
    \label{fig:mf_hybrid_relaxed_out}
    \end{subfigure}
    \newline
    \begin{subfigure}{.9\linewidth}
    \centering
    \vspace{1em}
    \includegraphics[width=.9\columnwidth]{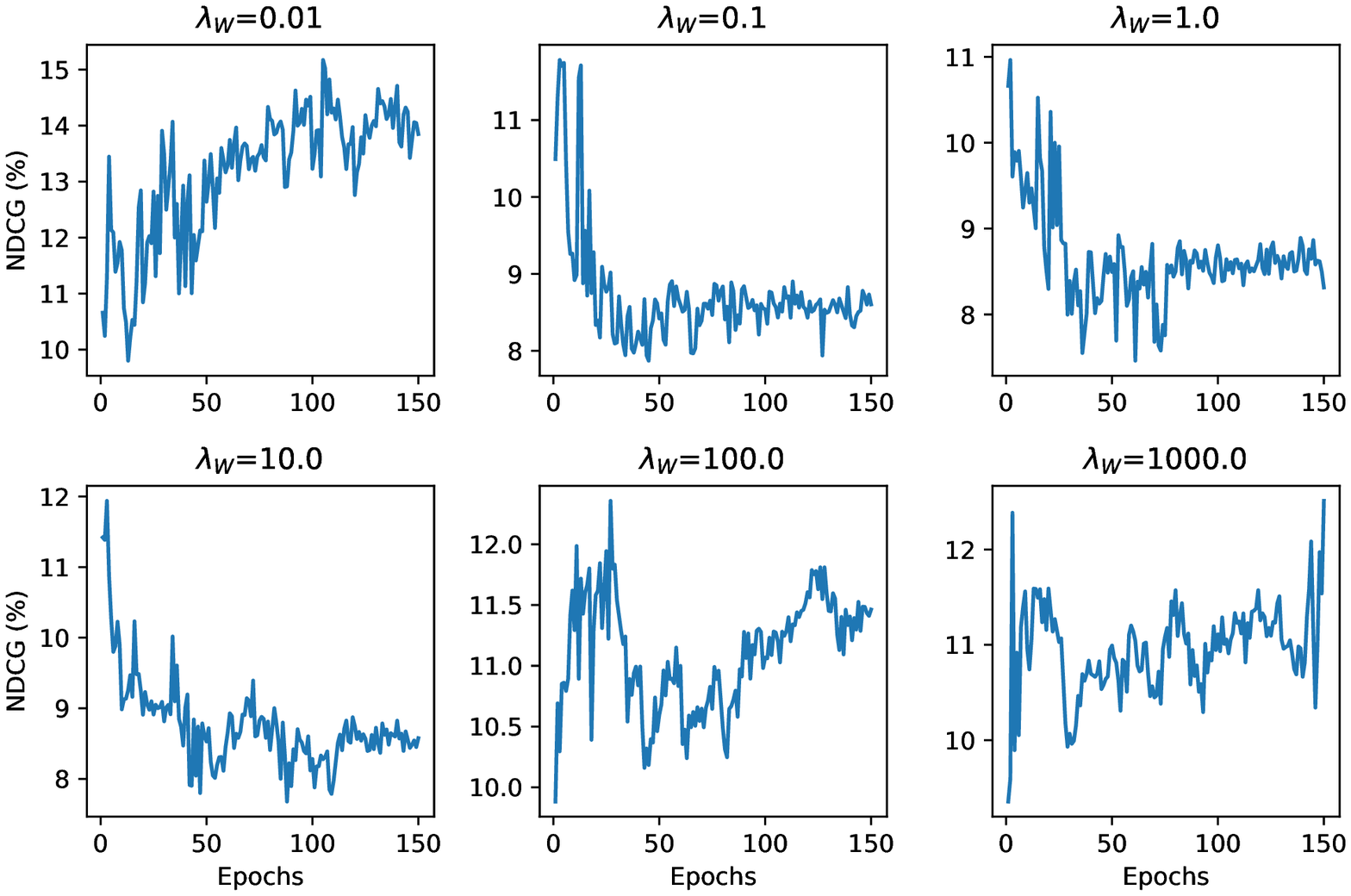}
    \vspace{-0.5em}
    \subcaption{Strict}
    \label{fig:mf_hybrid_strict_out}
    \end{subfigure}
    \caption{NDCG on the validation set for the MF-Hybrid algorithms (relaxed variant on the top, strict variant on the bottom) in the cold-start setting, for several values of the hyperparameters.}
    \label{fig:mf_hybrid_out}
\end{figure}

First, let us select the optimal hyperparameters for the MF-Hybrid algorithm (a similar experiment is conducted for optimizing DCB). The results in the warm-start scenario are presented in Fig.~\ref{fig:mf_hybrid_in}. It can be noted that the performance of MH-Hybrid-Relaxed does not strongly depend on these parameters (except for very large values of $\lambda_H$, where the performance decreases). The strict variant exhibits a less stable behavior overall and also yields worse results than its relaxed counterpart. A similar trend can be observed in the cold-start setting, whose results are presented in Fig.~\ref{fig:mf_hybrid_out}. More specifically, MF-Hybrid-Relaxed exhibits a better performance and a smoother (quasi-monotonic) behavior for large values of $\lambda_W$ and $\lambda_H$. This emphasizes the importance of a (relatively large) regularization on the user preference factor $\mathbf{W}$ in the cold-start setting.

\begin{table}[t]
    \centering
    \caption{Performance of the DCB and MF-Hybrid approaches (NDCG in $\%$). Note that in the warm-start setting, DCB is only presented in its strict variant since the relaxed one simply reduces to WMF.}
    \label{tab:mf}
    \begin{tabular}{llcc|cc} 
\hline\noalign{\smallskip}
     &  &  \multicolumn{2}{c|}{Warm-start} &   \multicolumn{2}{c}{Cold-start}  \\
     &   &  Relaxed   & Strict  & Relaxed   & Strict \\
\noalign{\smallskip}\hline\noalign{\smallskip}
    DCB  &  &  & $21.0$  & $17.5$  & $15.0$ \\
\noalign{\smallskip}\hline\noalign{\smallskip}
     & $N_{\text{gd}}=1$  &  $\textbf{28.2}$ & $\textbf{18.2}$ & $\textbf{19.4}$ & $\textbf{15.2}$ \\  
   MF-Hybrid & $N_{\text{gd}}=2$  &  $28.1$ & $13.8$ & $19.3$ & $13.8$ \\ 
    & $N_{\text{gd}}=5$  &  $28.0$ & $16.5$ & $19.3$ & $11.3$ \\ 
    & $N_{\text{gd}}=10$ &  $28.0$ & $15.9$ & $19.2$ & $11.3$ \\ 
\noalign{\smallskip}\hline
    \end{tabular}
\end{table}

Let us now investigate the impact of the parameter $N_{\text{gd}}$ on the performance of MF-Hybrid algorithms. Indeed, as can be read in Algorithms~\ref{al:MFhybrid_relaxed} and~\ref{al:MFhybrid_strict}, these approaches consist in alternating ALS updates for estimating $\mathbf{W}$ (and potentially $\mathbf{H}$ for the relaxed variant) and GD updates for estimating the deep content extractor's parameters. As such, the number of GD epochs performed in each iteration in between ALS updates is expected to have an impact on the overall performance. We test the MF-Hybrid algorithms with $N_{\text{gd}}=1$, $2$, $5$, and $10$ (the total number of epochs being the same), and we present the results in Table~\ref{tab:mf}.

For the relaxed variant, we observe that in both the warm- and cold-start scenarios, choosing a value of $N_{\text{gd}}$ between $1$ and $10$ yields overall similar results. However, a different behavior is observed for the strict variants, where the performance drops when increasing $N_{\text{gd}}$. Since the total number of epochs is fixed for a fair comparison, increasing $N_{\text{gd}}$ implies decreasing the total number of ALS updates, which might explain the decreasing performance in this case. The relaxed variants are therefore more promising since they exhibit less sensitivity to the number of GD updates.
Overall, updating the deep content feature extractor with a single epoch after updating the embeddings appears to be the optimal choice, as done e.g., in the CDL baseline, which corresponds to MF-Hybrid-Relaxed when $N_{\text{gd}}=1$.

Finally, we also report in Table~\ref{tab:mf} the performance of the DCB two-stage approach in its strict and relaxed variants, which completes the experimental results from Section~\ref{sec:results_overall}. We observe that the MF-Hybrid approaches overall outperform this baseline. In particular, in the cold-start scenario the performance improvement is more significant for the relaxed variant ($+1.9$ \%) than for its strict counterpart ($+0.2$ \%). This confirms the potential of jointly estimating the user/item factors and the deep content extractor, rather than relying on a two-stage approach. We also note that the relaxed variant outperforms its strict counterpart (for both DCB and MF-Hybrid approaches), which highlights the interest of using a more flexible system, especially in the cold-start recommendation scenario.

\subsection{Learning algorithms for the MF models}
\label{sec:results_hybvsuni}

In this experiment, we compare the learning strategies of the MF models, that is, whether they are trained using a combination of ALS and GD (MF-Hybrid) or a single GD algorithm (MF-Uni).

\begin{figure}[t]
\centering
    \begin{subfigure}{.99\linewidth}
    \hspace{0.8em} \includegraphics[width=.95\columnwidth]{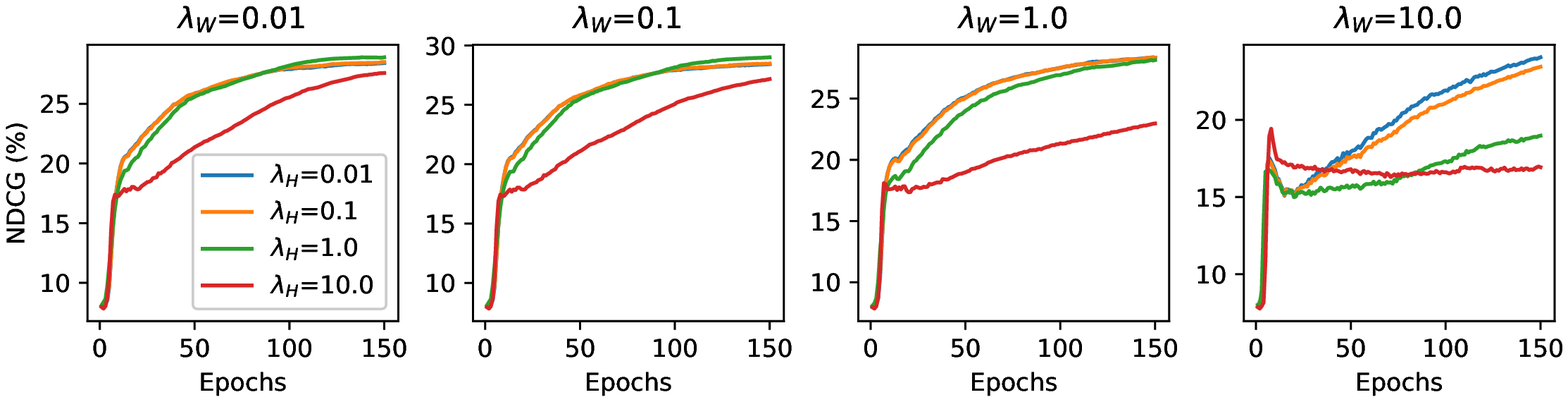}
    \vspace{-0.5em}
    \subcaption{Relaxed}
    \label{fig:mf_uni_relaxed_in}
    \end{subfigure}
    \newline
    \begin{subfigure}{.99\linewidth}
    \centering \includegraphics[width=.95\columnwidth]{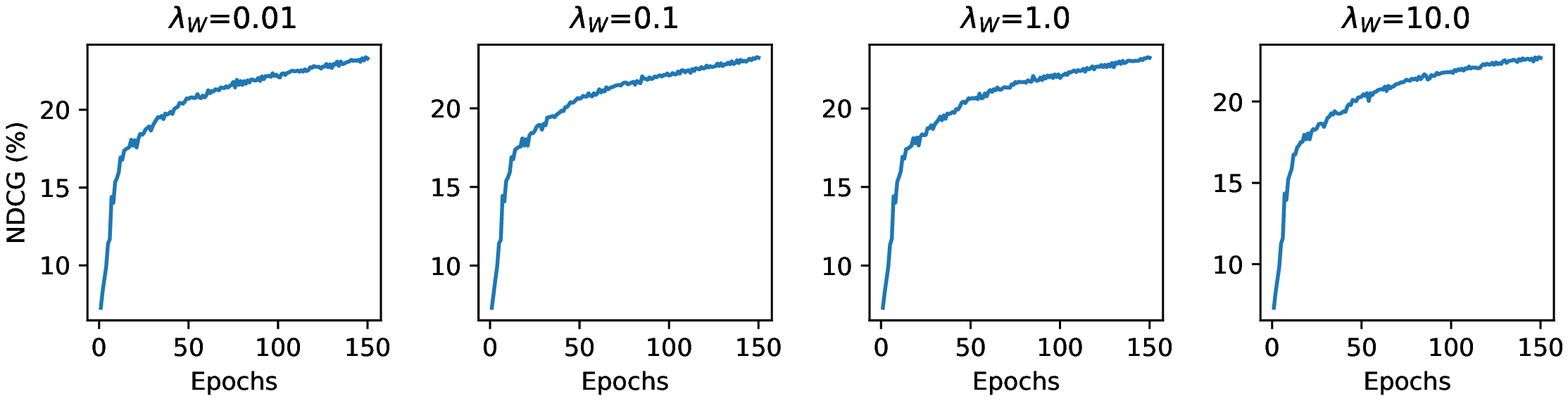}
    \vspace{-0.5em}
    \subcaption{Strict}
    \label{fig:mf_uni_strict_in}
    \end{subfigure}
    \caption{NDCG on the validation set for the MF-Uni algorithms (relaxed variant on the top, strict variant on the bottom) in the warm-start setting, for several values of the hyperparameters.}
    \label{fig:mf_uni_in}
\end{figure}

\begin{figure}[t]
\centering
    \begin{subfigure}{.99\linewidth}
    \hspace{0.8em} \includegraphics[width=.95\columnwidth]{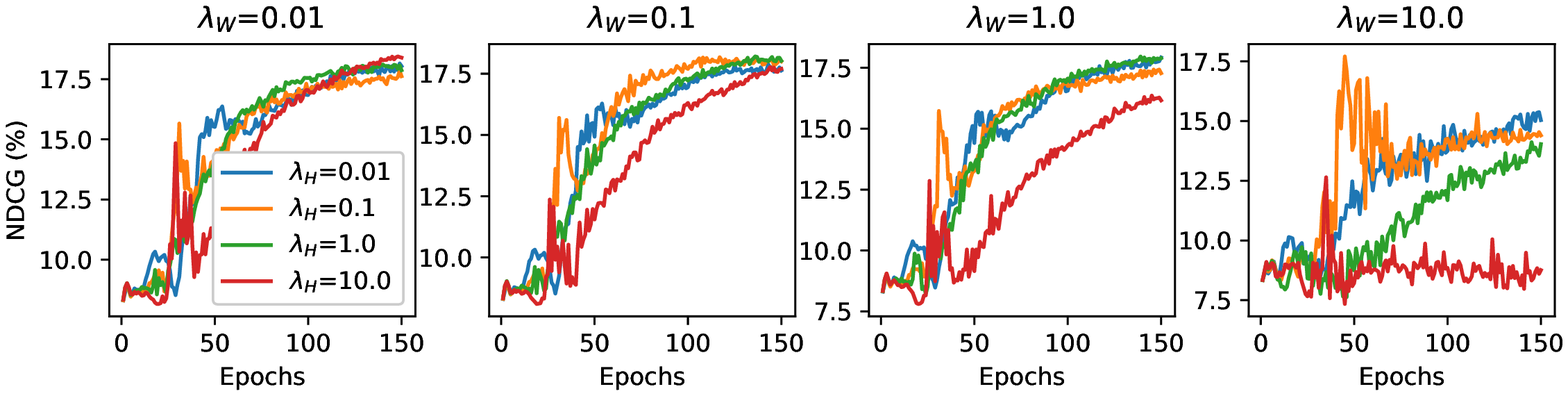}
    \vspace{-0.5em}
    \subcaption{Relaxed}
    \label{fig:mf_uni_relaxed_out}
    \end{subfigure}
    \newline
    \begin{subfigure}{.99\linewidth}
    \centering
    \includegraphics[width=.95\columnwidth]{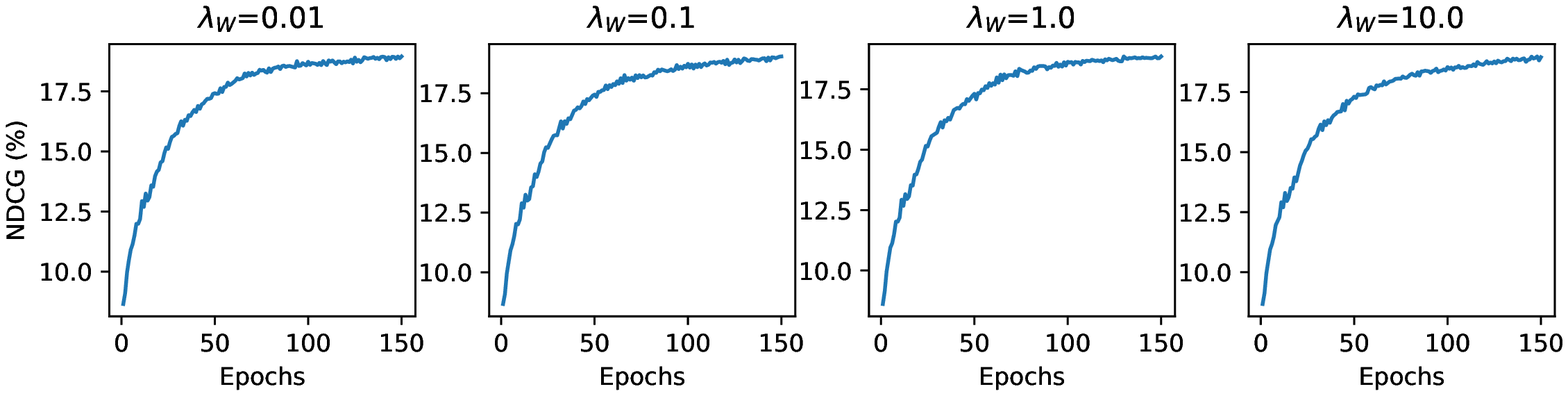}
    \vspace{-0.5em}
    \subcaption{Strict}
    \label{fig:mf_uni_strict_out}
    \end{subfigure}
    \caption{NDCG on the validation set for the MF-Uni algorithms (relaxed variant on the top, strict variant on the bottom) in the cold-start setting, for several values of the hyperparameters.}
    \label{fig:mf_uni_out}
\end{figure}

We first tune the hyperparameters for MF-Uni on the validation set and present the results in Fig.~\ref{fig:mf_uni_in} and~\ref{fig:mf_uni_out} for the warm- and cold-start scenarios, respectively. Interestingly, we observe than unlike MF-Hybrid, the performance in both variants (and in both scenarios) does not strongly depend on these values (except when very large values for  $\lambda_W$ and $\lambda_H$ are used, in which case the performance tends to decrease). This might explain why several previous studies incorporating the factors within the network did not specifically consider a regularization on the user preference matrix~\citep{He2017,Lee2018}.

\begin{table}[t]
    \centering
    \caption{Comparison between MF-Hybrid and MF-Uni.}
    \label{tab:mf_uni}
    \begin{tabular}{llcc|cc} 
\hline\noalign{\smallskip}
      & &  \multicolumn{2}{c|}{Warm-start} &   \multicolumn{2}{c}{Cold-start}  \\
      & &  NDCG ($\%$)   & Time (s)  & NDCG ($\%$) & Time (s) \\
\noalign{\smallskip}\hline\noalign{\smallskip}
    Relaxed   & MF-Hybrid &  $28.2$ & $\textbf{41}$  & $\textbf{19.4}$  & $\textbf{296}$ \\
              & MF-Uni    &  $\textbf{29.0}$ & $1525$  & $18.4$  & $1334$ \\
\noalign{\smallskip}\hline\noalign{\smallskip}
    Strict    & MF-Hybrid &  $18.2$ & $296$  & $15.2$  & $1048$ \\
              & MF-Uni    &  $23.4$ & $1334$  & $19.0$  & $1334$ \\
\noalign{\smallskip}\hline
    \end{tabular}
\end{table}

The performance of MF-Uni on the validation set is reported in Table~\ref{tab:mf_uni}, as well as the performance of MF-Hybrid with $N_{\text{gd}}=1$. We observe that in the cold-start scenario, the strict variant of MF-Uni significantly outperforms MF-Hybrid, but a different behavior is observed for the relaxed variant, where MF-Uni is outperformed by MF-Hybrid: this reveals the importance of choosing the learning algorithm in accordance with the underlying model. In the warm-start setting however, different conclusions are drawn: MF-Uni outperforms MF-Hybrid in both variants, but the best performing variants are the relaxed ones. This complements the results obtained in~\citep{Lee2018}, where DCUE (which is similar to MF-Uni-Strict) yielded a limited performance for a warm-start recommendation task: this model might have actually benefited from a more flexible formulation (MF-Uni-Relaxed), or might have revealed more potential for a cold-start recommendation task, where it is competitive with MH-Hybrid-Relaxed.

We also present in Table~\ref{tab:mf_uni} the total computational time for these methods. While MF-Uni-Relaxed allows for a slight improvement over MF-Hybrid-Relaxed in the warm-start setting, it is significantly more computationally demanding. The fastest method overall is MF-Hybrid-Relaxed: indeed, training the network in this approach only involves the target attributes $\mathbf{H}$, as opposed to its strict counterpart and to the MF-Uni approach, which involve the set of all binarized playcounts $\mathbf{R}$. Therefore, this experiment shows that when (part of) the model is tractable, leveraging closed-form updates in a flexible formulation is more interesting than treating all the factors as embedding layers in a deep network trained with GD, in terms of both computational time and performance. Incorporating user/item factors in the network then becomes interesting when the interaction model no longer allows for deriving closed-form updates, as will be shown in the next experiment.

\subsection{Impact of the interaction model in NCACF}
\label{sec:results_ncacf}

\begin{figure}[t]
\centering
    \begin{subfigure}{.9\linewidth}
    \centering \hspace{1.05em}
    \includegraphics[width=.8\columnwidth]{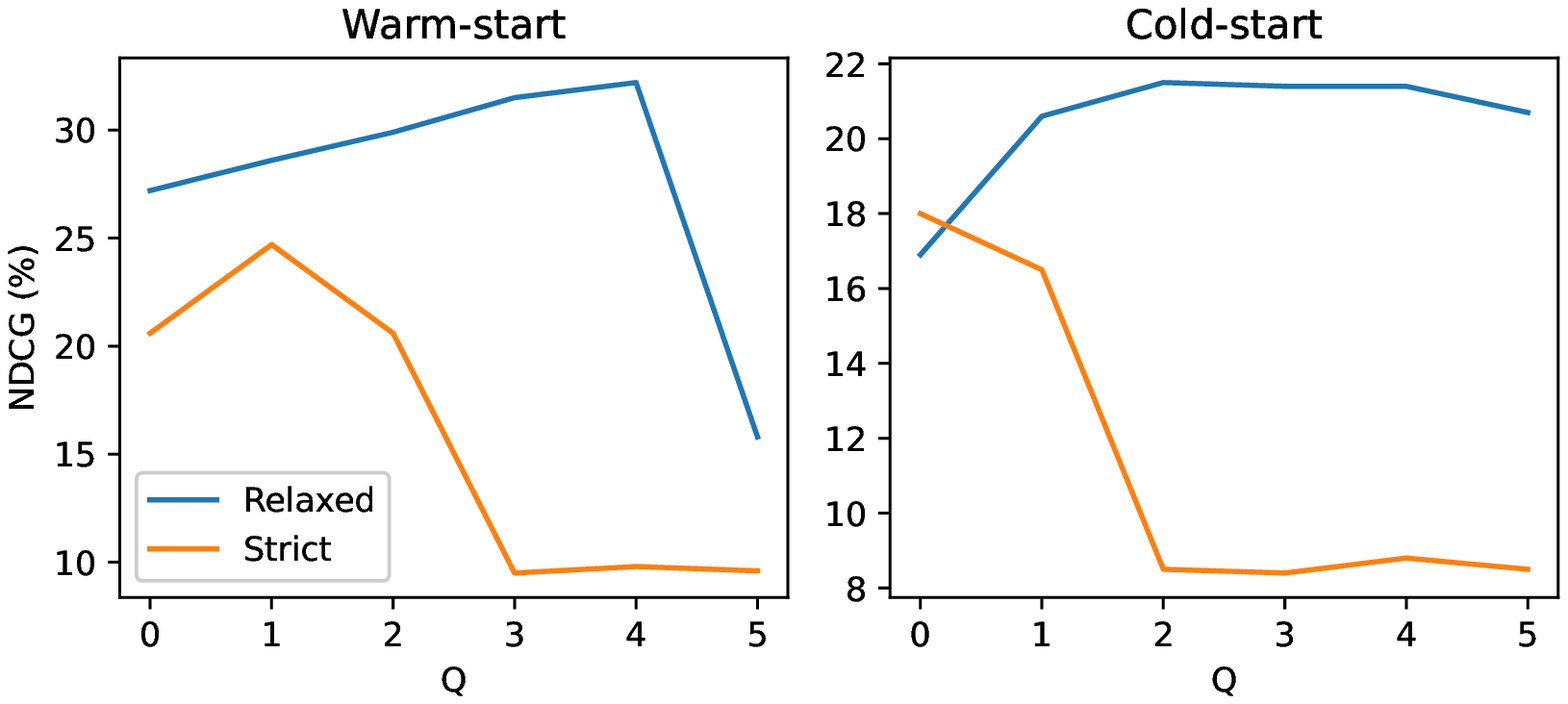}
    \vspace{-0.5em}
    \subcaption{Multiplication}
    \label{fig:ncacf_ndcg_mult}
    \end{subfigure}
    \newline
    \begin{subfigure}{.9\linewidth}
    \centering
    \includegraphics[width=.8\columnwidth]{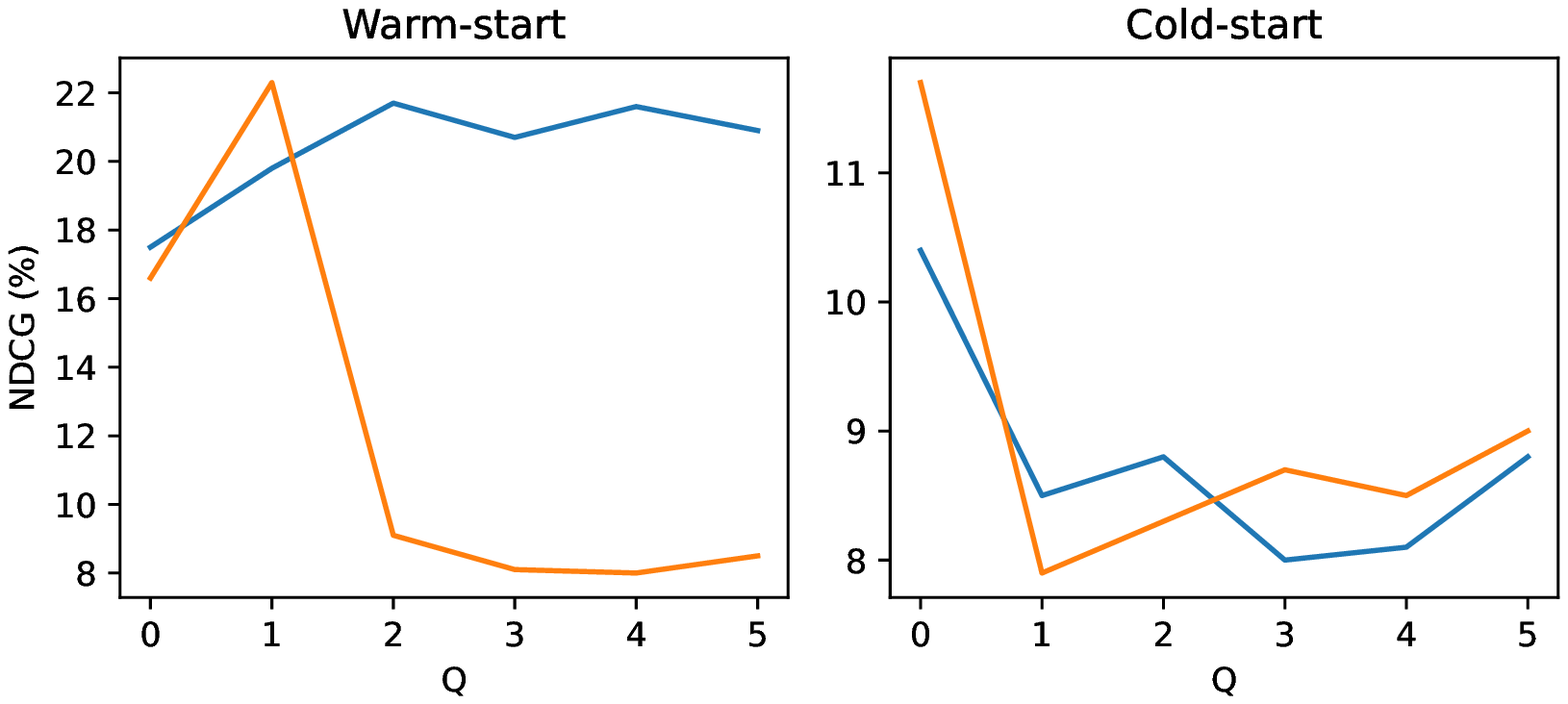}
    \vspace{-0.5em}
    \subcaption{Concatenation}
    \label{fig:ncacf_ndcg_conc}
    \end{subfigure}
    \caption{NDCG on the validation set for the NCACF model in warm-start (left) and cold-start (right) settings, for an interaction model based on multiplication (top) or concatenation (bottom) of the embeddings, for a variable amount of hidden layers $Q$.}
    \label{fig:test_ndcg_ncacf}
\end{figure}

In this experiment, we study the impact of the interaction model on the performance of NCACF. To save some computational time, we use the hyperparameters values obtained in the previous experiment. We consider two possible embedding combinations based on multiplication or concatenation, and a variable number of hidden layers $Q$ in the interaction network. The results are presented in Fig.~\ref{fig:test_ndcg_ncacf}.

First, we remark that overall the relaxed variants of NCACF benefit from using a deep interaction model, since we observe a performance increase (in most cases up to $Q=4$) when compared to a shallow interaction model ($Q=0$) (except for model using embedding concatenation in the cold-start scenario). We also observe that in most cases, the relaxed variants significantly outperform their strict counterparts, which seems more prone to overfitting when $Q$ increases. Nonetheless, in the warm-start scenario, the performance of NCACF-Strict (for both embedding multiplication and concatenation) remains superior to that of MF-Hybrid-Strict, which confirms the potential of fully learning a strict model with a single GD algorithm, provided a deep interaction model is used.

Interestingly, we remark that embedding multiplication outperforms concatenation by a large margin for all NCACF variants. Even though in the warm-setting embedding concatenation with $Q=1$ outperforms the baseline using a shallow interaction model, the improvement is more important when using embedding multiplication. This might be explained by the ability of the (non-linear) concatenation technique to learn complex interaction patterns between user and items for which some shared feedback is available, which in turns does not properly generalize to unseen items, where a more simple multiplicative model yields better performance. Overall, this result shows that several architectural design choices employed in ``pure" (i.e., warm-start) collaborative filtering techniques cannot be systematically exploited in the more challenging cold-start scenario, or when a deep content extractor is used. This somehow tempers conclusions from the literature, where embedding concatenation has been shown to perform better than multiplication in several studies addressing warm-start collaborative filtering~\citep{He2017,Chen2019}: while we observed that for NCF embedding multiplication and concatenation yield overall similar results, this does not generalize to its content-aware counterpart, even for warm-start recommendation.

This conclusion is reminiscent of recent work such as~\citep{Rendle2020}, where it is shown that a carefully tuned matrix factorization technique outperforms an NCF model using embeddings concatenation. The authors in~\citep{Rendle2020} also outline that a ``leave-one-out strategy", which is commonly employed for evaluating recommender systems~\citep{He2017,Chen2019}, is not appropriate for drawing general conclusions, notably when a different train/test splitting strategy is used, which is the case in our experiment as described in Section~\ref{sec:protocol_playcount}. Further investigation is then required to fully identify to what extent a concatenation of the embeddings is appropriate for addressing the cold-start problem.

\section{Conclusion}
\label{sec:conclu}

In this work, we introduced neural content-aware collaborative filtering, a unifying framework for cold-start recommendation which encompasses many methods from the literature. We proposed several variants of this model in order to study the impact of the learning strategy (two-stage vs. joint learning), training algorithm (hybrid vs. unified), and interaction model (shallow vs. deep). In particular, the proposed NCACF method fully leverages deep learning for modeling user/item interactions and extracting content information, and reaches state-of-the-art results for both warm- and cold-start recommendation on a large scale and highly sparse music dataset.

In future work, we will investigate alternative generative models that are better suited for the considered implicit feedback data. For instance, compound Poisson models~\citep{Gopalan2014,Gouvert2019} are appropriate for directly modeling the raw (non-binarized) playcounts and accounting for their over-dispersed nature. Alternatively, Bernoulli models~\citep{Bingham2009} are appropriate for modeling binarized playcounts, and yield alternative losses such as the binary cross entropy, which have shown superior performance than the quadratic loss~\citep{Chen2019}. We will also explore more refined sampling strategy~\citep{Chen2017, Tran2019} in order to reduce the computational time while keeping the performance high. Besides, our work can be extended to handle other modalities in order to fully exploit the available content such as artists biographies~\citep{Oramas2017} or musical tags~\citep{Liang2015,Zuo2016}, but also contextual data such as culture~\citep{Zangerle2020} or location~\citep{Gillhofer2015}. This work can also be combined with knowledge-enhanced recommenders~\citep{Wang2019,Wang2020,Li2020} in order to improve the interpretability of the item embeddings. Finally, alternative architectures could be exploited, such as convolutional networks that directly extract content features from the raw audio data~\citep{Lee2018} in an end-to-end fashion.

\bibliographystyle{spbasic}
\bibliography{references}

\end{document}